\documentclass{aa} 
\usepackage{graphicx,color}
\usepackage{txfonts}
\usepackage{natbib}
\bibpunct{(}{)}{;}{a}{}{,}
\def\ltapprox{\raise 2pt \hbox {$<$} \kern-1.1em \lower 5pt \hbox {$\approx$}}
\def\ltsim{\raise 2pt \hbox {$<$} \kern-1.1em \lower 4pt \hbox {$\sim$}}
\def\gtsim{\raise 2pt \hbox {$>$} \kern-1.1em \lower 4pt \hbox {$\sim$}}
\begin{document}
    \title{Total mass biases in X-ray galaxy clusters}

    \subtitle{}

    \author{R. Piffaretti
           \inst{1}
           \and
           R. Valdarnini\inst{1}\fnmsep
           }

    \offprints{R. Piffaretti}

    \institute{SISSA/ISAS, via Beirut 4, I-34014 Trieste, Italy
              }

    \date{Received ; accepted }


   \abstract
    {The exploitation of clusters of galaxies as cosmological probes relies on accurate measurements of their total gravitating mass. X-ray observations provide a powerful means of probing the total mass distribution in galaxy clusters, but might be affected by observational biases and rely on simplistic assumptions originating from our limited understanding of the intracluster medium physics.}
    {This paper is aimed at elucidating the reliability of X-ray total mass estimates in clusters of galaxies by properly disentangling various biases of both observational and physical origin.}
    {We use N-body/SPH simulation of a large sample of $\sim 100$ galaxy clusters and investigate total mass biases by comparing the mass reconstructed adopting an observational-like approach with the true mass in the simulations. X-ray surface brightness and temperature profiles extracted from the simulations are fitted with different models and adopting different radial fitting ranges in order to investigate modeling and extrapolation biases. Different theoretical definitions of gas temperature are used to investigate the effect of spectroscopic temperatures and a power ratio analysis of the surface brightness maps allows us to assess the dependence of the mass bias on cluster dynamical state. Moreover, we perform a study on the reliability of  hydrostatic and hydrodynamical equilibrium mass estimates using the full three-dimensional information in the simulation.}
    {A model with a low degree of sophistication such as the polytropic $\beta$-model can introduce, in comparison with a more adequate model, an additional mass underestimate of the order of $\sim 10 \%$ at $r_{\mathrm{500}}$ and $\sim 15 \%$ at $r_{\mathrm{200}}$. Underestimates due to extrapolation alone are at most of the order of $\sim 10 \%$ on average, but can be as large as  $\sim 50 \%$ for individual objects. Masses are on average biased lower for disturbed clusters than for relaxed ones and the scatter of the bias rapidly increases with increasingly disturbed dynamical state. The bias originating from spectroscopic temperatures alone is of the order of $10 \%$ at all radii for the whole numerical sample, but strongly depends on both dynamical state and cluster mass. From the full three dimensional information in the simulations we find that the hydrostatic equilibrium assumption yields masses underestimated by $\sim 10-15 \%$ and that masses computed by means of the hydrodynamical estimator are unbiased. Finally, we show that there is excellent agreement between our findings, results from similar analyses based on both Eulerian and Lagrangian simulations, and recent observational work based on the comparison between X-ray and gravitational lensing mass estimates.}
    {}


    \maketitle
%


\section{Introduction}
\label{intro.sec}
Galaxy clusters are the largest virialized structure known in the universe.
According to the hierarchical clustering model of structure formation 
they form by the gravitational collapse of the rare peaks of the primeval density 
field, on scales of the order of $\sim 10$ Mpc. Within this scenario their 
formation and evolution is a sensitive function of the cosmological matter
density parameter $\Omega_{\mathrm{m}}$ and the mass fluctuation amplitude 
$\sigma_\mathrm{8}$, where 
$\sigma_\mathrm{8}$ is the rms linear fluctuation on scales $8 \, h^{\mathrm{-1}}$ Mpc.

Measurements of their evolution rate can be used to 
asses the growth in mass 
of such structures, thereby providing a powerful method to constrain the 
geometry and matter content of the universe 
\citep[see][and references therein]{vo05}. 
Moreover, because of the spatial extent of the collapse scale, the cluster 
baryonic fraction $f_{\mathrm{b}}$ is expected to be close to the cosmic value $\Omega_{\mathrm{b}}/ \Omega_{\mathrm{m}}$.
Measurements of $f_{\mathrm{b}}$ at high redshifts  can be used to derive 
constraints on the equation of state of the dark energy \citep{hai01,maj04,al08}.

The importance of cluster of galaxies as cosmological probes will be further
strengthened with the upcoming high redshift surveys. This is of particular 
relevance in the new era of precision cosmology, in which studies of cluster
evolution will provide independent tests with which to compare constraints
on cosmological models extracted from observations of the cosmic background
radiation \citep[e.g.][]{spe07} and distance measurements of high redshift 
supernovae \citep[e.g.][]{ton03,rie04,rie07}.

From the scenario here outlined it follows that in order to use cluster 
of galaxies as cosmological probes it is crucial to accurately measure 
their baryonic and total gravitating mass. The methods used to derive cluster masses are  mainly based on the velocity dispersion of the optical galaxy populations
 \citep[e.g.][]{biv03,rin03}, observations of the X-ray emitting intracluster medium (ICM) \citep[e.g.][]{fin01,rei02,et02,arn05,vikhlinin2006} and on gravitational lensing \citep[e.g.][]{smi02,mahdavi08}.

Accurate mass estimates derived from X-ray data are based on the assumptions that both the total potential and the ICM distribution are spherically symmetric and that the ICM is in hydrostatic equilibrium in the cluster potential well. The latter assumption is usually justified by the fact that the estimated ICM sound crossing times are short when compared against cluster ages \citep{sar86}.

Under these assumptions the ICM is a faithful tracer of the underlying matter distribution and the total mass profile can be determined from the gas 
radial density and temperature profiles. The density profile is recovered from deprojecting of the surface brightness profile, as measured from X-ray flux maps, whereas knowledge of the temperature profile requires the availability of spatially resolved spectroscopy. High quality data taken by the present generation of X-ray telescopes,
 {\it Chandra} and {\it XMM-Newton}, allows for nearby clusters accurate 
measurements of these quantities out to a large fraction of the cluster virial 
radius \citep{ma98,vi05,pi05,san06,pra07,leccardi08}.

The reliability of cluster mass estimated through the X-ray method can be accurately studied using N-body/hydrodynamical simulations. The principal benefit over 
analytical methods is that the gas evolution can be treated self-consistently. 
The validity of the numerical approach is supported by X-ray observations, 
which show the existence of complex thermal structures  
and of merging activity \citep[see][for a review]{ma07}. 

Since early pioneering studies, hydrodynamical simulations have become a widely 
used tool to investigate cluster formation and evolution in different cosmological 
scenarios \citep[cf.][]{vo05}. 
The numerical resolution of the simulations and the modelization of the 
cluster gas physics  has been  improved over the years. The latter now incorporates radiative cooling \citep{yos00,lew20,per20,mua02,dav02},
metal enrichment of the ICM by supernovae and energy feedback 
\citep{kay03,tor03,valda03,bor04,kay04,kra05}.

The accuracy of cluster X-ray mass estimators has been tested by means of
N-body/hydrodynamical simulations in a variety of papers 
\citep{evr90,evr96,kay04,rasia06,kay07,nagai07}.
\cite{evr90} first pointed out the existence of a significant bias 
in the binding mass estimates when using the isothermal $\beta-$model.
\cite{evr96} confirmed that the source of this discrepancy is related 
to the isothermal and hydrostatic assumptions. 
The lack of validity of the hydrostatic assumption 
 is observationally motivated by optical and X-ray maps, which show
the existence of substructure with an ongoing merger activity, and
 is numerically supported by a number of authors \citep[e.g.][]{kay04,ras04,nagai07},
 who found that in the simulations the ICM is not perfectly in hydrostatic 
equilibrium. This implies the presence of 
residual gas bulk (laminar) and turbulent (random kinetic) motions, and leads to
 an underestimate of the masses because of additional non-thermal pressure support
 which is not accounted for by the  hydrostatic equilibrium equation.

With respect the isothermal $\beta-$model the modelization of the ICM has been
significantly improved \citep[e.g.][]{vikhlinin2006} with observational progresses, 
which showed the existence of temperature profiles declining with radius
\citep{de02,vi05,pi05,san06,pra07}. These features are well reproduced out of the core radii in hydrodynamical simulations which incorporate cooling and feedback \citep{mua02,kay03,tor03,valda03,bor04}. 

In order to properly assess the reliability of cluster X-ray mass estimators
it is however necessary to construct mock observations of simulated clusters 
which must reproduce spectroscopic measurements as expected from X-ray telescopes.
This is motivated by the presence of complex thermal structures in the ICM, which bias 
the (measured) spectral fit temperatures towards lower values than the average 
emission weighted cluster temperatures defined theoretically 
\citep{ma01,mazzotta04,valda06}. The dependence of X-ray mass estimators on 
spectral biases and other systematics
has been investigated through hydrodynamical simulations 
by a number of authors \citep{rasia06,kay07,nagai07,je07}.
\cite{nagai07} argued that mass estimates are biased low ($\sim 5-20 \%$)
even for clusters identified as relaxed. 

In order to properly disentangle observational biases which arise from spectroscopic measurements from those due to incomplete relaxation of the gas or from the ones caused by an inaccurate modeling of the radial profiles, it is however necessary to derive X-ray mass estimates from a large sample of simulated clusters.
This is the main goal of this paper, in which we apply different X-ray 
mass estimators to a large set of high-resolution hydrodynamical simulations of galaxy clusters. The physics of the gas includes radiative cooling, star formation, chemical enrichment and energy feedback. The sample comprises $\sim 100$ clusters, the size of the sample being a critical quantity in order to extract sub-samples large enough to give a meaningful statistics. 

We discuss the dependence of the mass bias at different radii upon the adopted 
analytical models and the chosen radial range used to perform the fits of the 
profiles,  the cluster dynamical state as well as the impact on the mass bias 
which follows from the use of spectral temperatures. As a statistical indicator to quantify the cluster dynamical state through 
the analysis of X-ray maps we adopt the power ratio method \citep[see][and references therein]{je05}.
We also investigate the limit of applicability of the dynamical equilibrium 
equation when used to recover the cluster true masses in the presence of 
significant non-thermal gas pressure. 

The paper is organized as follows. In Sect. \ref{simulations.sec} we present the procedure for simulating the cluster sample. In Sect. \ref{observables.sec} we describe how we generate and analyze the synthetic X-ray observations that are used in Sect. \ref{mass.sec} to recover the total mass distribution. In Sect. \ref{masssim.sec} we investigate the reliability of the hydrostatic and hydrodynamical mass estimators from the full three-dimensional information provided by the simulations. Finally, we discuss our main results and present our conclusions in Sect. \ref{conclusions.sec} .

 \section{N-body/SPH simulations}
\label{simulations.sec}
The considered cosmological model assumes a flat CDM universe, with matter density 
parameter $\Omega_\mathrm{m}=0.3$, $\Omega_\mathrm{\Lambda}=0.7$, 
$\Omega_\mathrm{b}=0.0486$ and $h=0.7$ is the value of the Hubble constant in units of 100 km $\mathrm{sec}^{-1} \mathrm{Mpc}^{-1}$. The power spectrum has been normalized to 
$\sigma_\mathrm{8}=0.9$ on a $8 \, h^{-1}$ Mpc scale at the present epoch and the primeval spectral index $n$ is set to $1$.

The simulation ensemble of galaxy clusters is constructed according to the following procedures. A low-resolution N-body run 
is first performed starting from an initial reshift $z_\mathrm{i}=10$
in a box of comoving size $L$, using a P3M code with $N_\mathrm{p}$ particles. Clusters of galaxies are identified at $z=0$ using a friends-of-friends 
(FoF) algorithm. The virial mass and radius are related by
   $M_\mathrm{vir}= (4 \pi/3) \, \Omega_\mathrm{m} \, \rho_\mathrm{c} \, \Delta_\mathrm{c} \, r_\mathrm{vir}^3$,
where $\Delta_\mathrm{c} \simeq 187 \, \Omega_\mathrm{m}^{-0.55}$ for a flat cosmology and $\rho_\mathrm{c}$ is the critical density. This is defined as $\rho_\mathrm{c}(z)=3 H(z)^{2} /8 \pi G $, where $H(z)^2=H_\mathrm{0}^2 E(z)^2$ and
$E(z)^2=\Omega_\mathrm{m} (1+z)^3 + \Omega_\mathrm{\Lambda}$. In general, the fiducial radius $r_{\mathrm{\Delta}}$ is defined such that
 the average of the total density within that radius is $\Delta$ times the critical density, i.e. $M^{}(<r_{\mathrm{\Delta}})= {4 \pi r_{\mathrm{\Delta}}^3} 
\, \Delta \,  \rho_\mathrm{c}(z) /3$.

The simulation ensemble is constructed by combining two 
distinct samples S1 and S2. Table \ref{tab:sims} lists the most relevant simulation 
parameters of the two samples S1 and S2.
\begin{table}[ht]   
\caption{
\label{tab:sims} 
Main parameters of the two cluster samples S1 and S2: comoving box size L of the cosmological $N-$body runs, number of corresponding particles $N_p$, number $N_{cl}$ of the most massive clusters extracted from the sample at $z=0$, and virial mass $M_{vir}$ of the most (least) massive cluster of sample S1(S2).}
\centering                            
\begin{tabular}{|ccccc|}        
\hline\hline                 
Sample & L & $N_p$ & $N_{cl}$ & $M_{vir}$ \\  
& $[h^{-1} \mathrm{Mpc}]$& &  &$[10^{14} \, h^{-1} M_{\odot}]$ \\
\hline                       
S1  & $400$ & $168^3$ & $33$	& $15 $  \\
S2  & $200$ & $84^3$ & $120$	& $1.5$  \\
\hline                                  
\end{tabular}
\end{table}

A sample is constructed by extracting at $z=0$ the $N_{cl}$ most massive 
clusters which are found in a cosmological run of size $L$ and
 $N_{p}$ particles. 
For sample S1  the box size and the number of particles are twice and
eight times those of sample S2, respectively.
The value of $N_{cl}$ for sample S1 is  
chosen such that the last cluster of the sample has its 
virial mass above that of the first cluster of sample S2. The mass range for 
the virial masses of the combined samples spans
a decade. The random realization of the initial density perturbations 
are different in the two cosmological simulations.

After completition of the cluster selection, the clusters of the ensemble
are resimulated individually using high-resolution hydrodynamic simulations in 
physical coordinates. The initial conditions of each hydrodynamic simulation are set as follows. The particles of a cluster lying  at $z=0$ within a distance $r_\mathrm{vir}$ from the cluster center are tagged and located back at an initial redshift $z_\mathrm{in}=49$ and a cubic region of size $L_\mathrm{c} \simeq 25-50 \, \mathrm{Mpc} \propto M_\mathrm{vir} ^{1/3}$, which contains these particles, is placed at the cluster center.
The original Fourier modes of the cosmological simulation are then used to perturb 
the positions of a uniform lattice of $N_{L}=51^3$ gas particles set inside the 
cube and high frequency waves are added to the original modes to sample the new 
Nyquist frequency. The positions of dark matter particles are found similarly. 
Particles with perturbed positions  which lie within a inner sphere of radius 
$L_\mathrm{c}/2$ from the cluster centre are kept for the hydrodynamical 
simulations. For these particles the masses are set in proportion to
$\Omega_\mathrm{b}$  and $(\Omega_\mathrm{m} - \Omega_\mathrm{b})$ 
for gas and dark matter particles, respectively.
Tidal fields are modeled by adding to the inner particles an external 
shell of low-resolution dark matter particles. The shell has an outer radius 
$L_\mathrm{c}$ and the mass of a particle is $8$ times the sum of the masses of a 
gas and dark matter particle of the inner region.

The hydrodynamic simulations are run using a multistep TREESPH code 
 in which the gas entropy is explicitly conserved \citep{go91}.
The simulations contain $\simeq 70,000$ gas and dark matter particles 
in the inner region and a similar value of low-resolution dark matter particles 
in the outer shell. 
The mass of the gas particles ranges from 
$m_\mathrm{g} \simeq 5 \times 10^9  M_{\odot}$ for the most massive cluster of 
the ensemble, down to $m_\mathrm{g} \simeq 6 \times  10^8 M_{\odot}$ for the 
least massive cluster. This mass resolution can be considered adequate for the 
present purposes, as suggested from analyzing the stability of gas profiles of 
simulated clusters \citep{valda02}.
The  gravitational softening parameter of gas particles is set 
to $\varepsilon_\mathrm{g} =25$ kpc and 15 kpc, for clusters of sample S1 and S2, 
respectively. For the dark matter particles  the softening is rescaled 
according to  $\varepsilon_\mathrm{i} \propto m_\mathrm{i}^{1/3}$, where 
$m_\mathrm{i}$ is the mass of the particle $i$. The softenings are comoving out 
to $z=20$, after which are kept fixed in physical coordinates.

The physical modeling of the gas includes radiative cooling, which 
depends on the gas temperature and metallicity. Cold gas in high density regions is 
subject to star formation and gas particles are eligible to form star particles. 
At each timestep gas particles neighboring a star particle are heated by 
supernova (SN) explosions of type II and Ia. The gas is also metal enriched through 
SN explosions. The energy and metal feedback are calculated according to the 
stellar lifetime and initial mass function. A detailed description of the  feedback
 recipes is given in \citet{valda03}. The hydrodynamic variables of a simulated 
cluster are stored at run time at various redshifts in the interval from $z=2$ down to $z=0$.

\section{Simulation and analysis of X-ray observations}
\label{observables.sec}
At  $z=0$ we extract a total of 153 clusters. For this sample we compute spectroscopic-like global temperatures and select objects with  $T_\mathrm{sl}(<r_\mathrm{500}) \geq 2$ keV (see Sect. \ref{tprof.sec} below for the definition of $T_\mathrm{sl}$). This temperature selection is adopted because when generating spectroscopic temperature profiles we use the approximation derived by \cite{mazzotta04}, which was developed for continuum-dominated spectra and is therefore not accurate for low temperature systems. Our final sample comprises $\sim$ 100 temperature-selected clusters spanning a mass range of $8.2 \times  10^{13} \, h^{-1 }M_{\odot}   \lesssim M(<r_\mathrm{200}) \lesssim 1.2 \times  10^{15} \, h^{-1 }M_{\odot}$.

For each simulated cluster we generate three independent observations by considering three orthogonal projections, thus obtaining a total of $\sim$ 300 observed objects. In following sections we describe how we generate and analyze these mock X-ray observations. Our analysis may be viewed as complementary to those presented in \cite{rasia06} and
 \cite{nagai07}.  The main difference being the large size of the sample 
 used here, which allows to draw results of much higher statistical 
significance.

\subsection{ X-ray maps }
\label{sec:map}
Simulated surface brightness maps are obtained by first choosing a line of sight 
and then locating the origin at the cluster center in the plane 
orthogonal to the line of sight, the latter is defined as the location where the 
gas density reaches its maximum value. Throughout the paper we assume that 
this position also coincides with the peak of the X-ray emission.

For a given line of sight and redshift $z$ the X-ray surface brightness is defined as:
\begin{equation}
\label{sig}
S_X (x,y) = \frac{1}{4 \pi (1+z)^4}\int_{-\infty}^{\infty} \varepsilon_X d l ~ ,
\end{equation}
where the integral is along the line of sight, $x$ and $y$ are cartesian coordinates on the chosen plane  and $\varepsilon_\mathrm{X}$ is the X-ray emissivity. The latter quantity can be calculated as 
$\varepsilon_\mathrm{X} = \Lambda \, \rho_\mathrm{g}^2({\bf x})$, 
which differs from the true emissivity aside from a constant factor.
Here  $\rho_\mathrm{g} ({\bf x})$ is the gas density, 
$\Lambda = \Lambda(T,Z,{\bf x},E_1,E_2)$ 
is the cooling function, and $T$ and $Z$ are the gas temperature and metallicity, respectively. The quantities $E_1$ and $E_2$ define the energy band [$E_1$-$E_2$] used in the X-ray flux measurement. Because of the Lagrangian nature of SPH simulations, the emissivity 
$\varepsilon_\mathrm{X}( {\bf x})$ is expressed as a summation over particles:

\begin{equation}
\label{eps}
\varepsilon_X ({\bf  x}) = \sum_j \frac{m_j} {\rho_{g,j}}\Lambda(T_j,Z_j,{\bf x},E_1,E_2) \rho_{g,j}^2 ({\bf x})  W({\bf x-x_j},h_j)~,
\end{equation}
   here the subscript $j$ denotes the values of the quantity at the position of the 
particle $j$,  $W$ is the SPH smoothing kernel, and $h_j$ the smoothing length
of particle $j$. From this expression for  $\varepsilon_\mathrm{X} ({\bf  x})$  it is 
possible to evaluate the X--ray surface brightness  $S_\mathrm{X}(x,y)$ by
performing the integral (\ref{sig})  along the line of sight axis, $\tilde{z}$. 
For our purposes the  integral (\ref{sig})  is replaced by

\begin{equation}
\label{sigb}
S_X (x,y) = \frac{1}{4 \pi (1+z)^4} \varepsilon_X^{2D}(x,y),
\end{equation}
where $\varepsilon^{2D}$ is defined according to eq. (\ref{eps}) in 
which  the kernel is the 2D Gaussian kernel $W_{2D}=exp[-(\sqrt{ x^2+y^2}/h_j^G)^2]/
\pi h_{j}^{G \;2}$ and $h_j^G \simeq 0.82 h_j$.

For each simulated object we generate three surface brightness maps by choosing three orthogonal projections. 
Motivated by observations we evaluate the maps  in the soft energy band [0.5-2] keV. 
Since particles with temperatures below $10^{5} \, K$ do not contribute to the gas
 emissivity in this energy band, they are not taken into account in the 
computation.

Cold and dense ICM clumps can affect total mass reconstructions because they produce pronounced irregular features in azimuthally averaged brightness and temperature profiles, and introduce a systematic bias in the spectral temperature determination. We therefore remove small scale clumps from the constructed flux images by performing a masking procedure as done with real X-ray observations. This is accomplished by generating [0.5-2] keV 
maps on  a $1024 \times 1024$ grid of size $2.1 \times r_{200}$. The pixel size of the maps is $\sim 5$ and $2$ kpc for the most and least massive clusters in our sample, respectively. Pixel sizes are similar to those adopted in observations and sufficiently small as to allow a complete identification of all small-scale clumps resolved in our simulations. Clumps are detected and removed following the procedure implemented by \cite{vi98}: images are decomposed into wavelets of pixel scale $2^n$, with $n=0,1,\ldots 6$, with significance threshold set to 5 in units of the rms level.
Particles which lie in those pixels which are part of a statistically 
significant structure on a given scale are tagged and identified as part of a clump.
These particles are then removed from all the SPH summations.
In three dimensions clumps are identified by those particles which are part
 of a two dimensional clump in any of the planes. The wavelets algorithm commonly detects also the central peak of the surface brightness distribution. In order to avoid the unwanted masking of the cluster emission in the central region we exclude the inner $0.2 \times r_{200}$ circular aperture from our masking procedure. In addition, we do not impose any limiting flux $f_X$ for the detection of small
scale clumps. We motivate this choice with the findings 
that setting $f_X$ to $\sim 3 \times 10^{-15} erg \, /cm^{-2} \, s^{-1}$, as in \cite{nagai07}, has fully negligible effects. Finally, we compute the amount of gas mass removed by the masking procedure and find that even in the most clumpy clusters it is less than a few percent of the total gas mass. If not stated explicitly, all the results presented in the following are derived by including the masking procedure.

\subsection{Power ratios}
\label{powerratios.sec}

Total mass determinations from X-ray observations can be heavily affected by the
  cluster dynamical state since the method relies on the assumption of hydrostatic
  equilibrium. The dynamical state of clusters is related to the amount of substructure
  present in their X-ray surface brightness
distribution \citep[e.g.][]{ri92,evrard93} and various 
statistical measures have been proposed to quantify cluster substructure 
\citep[e.g.][and references therein]{bu02}.

In this work we adopt the power ratio method \citep{bu95} as a statistical indicator of the cluster dynamical state, since it 
is widely used to study cluster X-ray morphologies \citep[e.g.][]{bu96}. This method is expect to be statistically significant when applied to a large cluster sample such as the one studied here. It is in fact unaffected only by mergers along the line of sight which rarely occur.

According to the power ratio method, the X-ray surface brightness map 
$S_\mathrm{X}(\rho,\varphi)$, where $(\rho,\varphi)$ are the conventional polar coordinates, is the source term of the pseudo potential 
$\Psi(\rho,\varphi)$ which satisfies the 2-D Poisson equation. The pseudo potential 
is expanded into plane harmonics and the $m$-th coefficients of the expansion are given by :

\begin{equation}
\alpha_m = \int_{R^{'}\leq R_{ap}}  d ^2 x^{'}  S_X(\vec x^{'}) {R^{'}}^m
\cos(m\varphi^{'}),
\label{am}
\end{equation}
\begin{equation}
\beta_m = \int_{R^{'}\leq R_{ap}}  d ^2 x^{'}  S_X(\vec x^{'}) {R^{'}}^m
\sin(m\varphi^{'}),
\label{bm}
\end{equation}

where $\vec x^{'}=(\rho,\varphi)$ and the integration is over a circular region with aperture radius $R_{ap}$.
The  $m$-th power ratio is then defined as
\begin{equation}
\Pi_m (R_{ap}) = \log_{10} (P_m/P_0)~,
\end{equation}
where
\begin{eqnarray}
P_m (R_{ap})& =& {1 \over 2 m^2} (\alpha_m^2 + \beta_m^2) ~~m>0,\\
P_0 &= &[ \alpha_0 \ln(R_{ap}/{\rm kpc}) ]^2.
\end{eqnarray}

The power ratios $\Pi_\mathrm{m}(R_\mathrm{ap})$ are then indicators of the amount of structure 
present 
on the scale of the aperture radius $R_\mathrm{ap}$. The values of $P_\mathrm{m}$ depend on the choice of the coordinate system.
For a fully relaxed configuration 
$\Pi_\mathrm{m}\rightarrow - \infty$. As $\Pi_3$ indicates asymmetric distributions, we 
adopt it here as a measure of the amount of substructure present in a cluster.

For each of the $\sim 300$ [0.5-2] keV surface brightness maps (see Sect. \ref{sec:map}) we set the origin of coordinates at the peak of the X-ray emission. In addition to being fully consistent with the procedure employed in the computation of azimuthally averaged brightness and temperature profiles, this choice of coordinate system enables us to detect $P_\mathrm{3}$ values different from zero even in the case of bimodal clusters with nearly equal size components. For the same configuration all the odd moments would vanish if the coordinate system was the frequently adopted centroid frame, which is the coordinate system defined such that $P_\mathrm{1}=0$ \citep{bu95}. We utilize unmasked maps, since the goal of the power ratio analysis is to measure substructure, and for each one we compute $\Pi_3(R_\mathrm{ap})$ in correspondence of the same radii at which we evaluate the mass biases ($r_{\mathrm{2500}}$, $r_{\mathrm{500}}$, and $r_{\mathrm{200}}$, see Sect. \ref{mass.sec}). The various integrals involved in 
the computations are performed according to the SPH prescription. For the simulated clusters of our sample, the values of $\Pi_3$ we found lie in the range between $\sim -(6-4)$ for a strongly asymmetric distribution and $\sim -(12-10)$  for a cluster with a relaxed configuration.

In order to perform a statistical analysis of mass determination biases as a function of substructure we extract four sub-samples in the following way. For a given overdensity, we first construct the cumulative distribution of the sample values of $\Pi_\mathrm{3}$. Then we identify the synthetically observed clusters in four sub-samples ($\Pi_3$ classes, hereafter): first quartile (below $25\%$), second
quartile (between $25\%$ and $50\%$), third quartile (between $50\%$ and $75\%$), and forth quartile (above $75\%$). In the following these four sub-samples will be referred to as $q_\mathrm{1}$, $q_\mathrm{2}$, $q_\mathrm{3}$, and $q_\mathrm{4}$ class, respectively. By construction the most relaxed clusters belong to the $q_\mathrm{1}$ class and the most disturbed ones to the $q_\mathrm{4}$ class.

\subsection{Computation an modeling of surface brightness profiles}
\label{sx.sec}
The azimuthally averaged surface brightness profiles $S_{\mathrm{X}}(r)$ are 
computed from the [0.5-2] keV band surface brightness maps as follows. Each map is binned using a grid with circular geometry and 
coordinates ($\tilde {\rho}, \tilde{\phi}$), the origin is set at the position of the
X-ray emission peak. The grid points are uniformly spaced, linearly in the angular
coordinate, and logarithmically in the radial coordinate.
The range of the spatial coordinates is between $2 \times 10^{-4}$ and $1.5$
in units of $r_{200}$.
There are $N_{\tilde {\rho}}=140, N_{\tilde{\phi}}=20$ points
in the coordinate intervals. The number and spacing of the radial grid points as 
chosen such that the surface brightness binning in the radial range used in the 
fits (see Sect. \ref{mass.sec} below) is similar to the one adopted in the analysis
  of real X-ray data. The value of $S_\mathrm{X}(r)$ at a given projected radius $r$ is given
  by averaging
over the azimuthal values of the annulus. These values are also used to define
   a surface brightness dispersion, which is later used to define a weight 
in correspondence of $S_\mathrm{X}(r)$ when the profile is fitted.

In order to explore the influence of the models adopted in the fitting of surface brightness radial profiles on total mass estimates, we adopt two different models. In the simplest and most widely adopted procedure used in the determination of the total gravitational mass of clusters the surface brightness profile is modeled by a single $\beta$-model \citep{cavaliere76}, i.e.
\begin{equation}
S_X(r) = S_0 \left( 1+ \frac{r^2}{r^2_{\rm c}} \right)^{-3\beta + 0.5}.
\label{1beta-sb.eq} 
\end{equation}
The single $\beta$-model is convenient because it is simple to deproject it and obtain the ICM density profile needed to estimate the total mass from the hydrostatic equilibrium.

As a second parametrization we adopt an extended $\beta$-model in which the profiles are modeled as in \cite{vikhlinin2006}. The radial dependence of $\rho_\mathrm{g}(r)$ is given by the functional form:
\begin{equation}
\label{rho}
\rho_g^2(r)=\rho_{g,0}^2\frac{(r/r_c)^{-\alpha}}{(1+r^2/r_c^2)^{3\beta-\alpha/2}}
\frac{1}{(1+r^{\gamma}/r_s^{\gamma})^{\varepsilon/\gamma}},
\end{equation}
where the additional parameters with respect the standard $\beta-$model allow a much more accurate modeling of the density slope variations. The component present in in Eq. (\ref{rho}) which takes into account for a central surface brightness excess is neglected (i.e. we set  $\alpha=0$), since, as discussed below, we are excising the very central regions in the fitting procedure. The model parameters are recovered by fitting the projection of Eq. (\ref{rho}) along the line of sight against the surface brightness profiles in the [0.5-2] keV energy band. Following the suggestions of \cite{vikhlinin2006} we set in Eq. (\ref{rho}) $\gamma=3$.

Notice that for both parametrizations the total mass derived from the hydrostatic equilibrium equation (Eq. (\ref{mass.eq}) below) does not depend on the central gas density value.

As commonly found in simulated clusters, the gas properties in the central regions of our simulated objects are in disagreement with observations. The pronounced steepening of the gas density towards the center in low temperature systems and the sudden temperature drop in the very inner regions are in fact not observed in real clusters. We therefore 
exclude the inner region within $r_\mathrm{low}=0.1 \times r_\mathrm{200}$ from the surface brightness fits. We emphasize that the precise value for $r_\mathrm{low}$ is not relevant, since our goal is to quantify the bias in the mass reconstruction at much larger radii.

\subsection{Computation an modeling of temperature profiles}
\label{tprof.sec}
From the three-dimensional gas temperature distribution $T^{3D}(\vec x )$ measured in the simulated clusters we compute both three- and two-dimensional (projected) radial profiles.

A three-dimensional radial temperature profile can be defined as 

\begin{equation}
T_{W}^{3D}(r)= \frac {\int_{\Delta V} T^{3D}(\vec x ) {\it W} d^3x } { \int_{\Delta V} {\it W} d^3x},
\label{tr}
\end{equation}

where $W$ is a weight function and the volume integral is over a  
spherical shell of thickness $\Delta r$ located at distance $r$ from the cluster center. 
Common choices for $\it W$ are the gas 
density (mass-weighted temperatures, ${\it W}=\rho_\mathrm{g}$) and 
the X-ray emissivity  (emission-weighted temperatures, ${\it W}=\varepsilon_\mathrm{X}$). In this work we consider mass-weighted temperatures $T_\mathrm{mw}$ and the weight function ${\it W}=\rho^2_\mathrm{g} T^{-3/4}$,
 which defines  the corresponding spectroscopic-like temperature $T_\mathrm{sl}$. \cite{mazzotta04} found that in the continuum regime ($T \gtrsim 2$keV) this choice of the weight function provides accurate approximations of spectroscopic temperatures obtained from X-ray observations (i.e. derived by fitting an integrated spectrum with a single-temperature model). A more sophisticated method would be based on the temperature determinations from either projected spectra (in the case of projected temperature profiles) or from deprojected ones (in the case of deprojected temperature profiles) computed from the simulation outputs. Given the size of our sample and being this procedure computationally expensive, we adopt spectroscopic-like temperature profiles. 
 
 For the cluster under consideration we evaluate the weighted 
 temperature profiles by replacing the integrals in Eq. (\ref{tr}),
according to the SPH scheme, by a summation over  gas particles. The value of  the radial temperature profile $ T_{\it W}^{3D}(r)$
 in correspondence of each shell is defined by averaging over a set of 
$(\theta,\phi)=40 x 40 $
grid points uniformly spaced in angular coordinates. The radial coordinate of the 
shells is binned as in the computation of the surface brightness. For each object we therefore obtain two types of three-dimensional temperature profiles: a mass-weighted profile and a spectroscopic-like one.

The projected temperature profiles $T_{W}^{2D}(r)$ are computed in the same way as the three-dimensional profiles, but by considering a cylindrical geometry instead of a spherical one. Furthermore, only spectroscopic-like profiles are computed, since 
mass-weighted projected profiles are not meaningful in this case. For each cluster we compute three projected spectroscopic-like temperature profiles by consistently considering the same projections used in the computation of surface brightness profiles. 

We compare projected to three-dimensional spectroscopic-like profiles and find that in the radial range [0.1-1] $r_{\mathrm{200}}$  the difference is generally small. The agreement between the two types of profiles steadily improves from disturbed to relaxed clusters and the difference is fully negligible for very relaxed clusters.

In the computation of temperatures, in accord with the considered energy bands, we neglect  in the SPH summations particles with temperatures below $10^{5}$ K. For consistency, 
we adopt the same temperature cut-off in the computation of 
mass-weighted temperatures. Moreover, for the spectral weighting we restrict the summation to those gas particles for which $kT > 0.5$ keV. 

It is crucial to remark that spectroscopic-like temperatures are good approximations of the spectroscopic temperatures as would be estimated from X-ray observations, whereas mass-weighted temperatures are a much more accurate indicator of the total binding mass, since they follow the virial relationship.

As for the brightness profiles, we use two different models to fit temperature profiles and exclude the inner aperture within $r_\mathrm{low}=0.1 \times r_\mathrm{200}$ from the fits. In the simplest case the gas temperature profile is modeled using the commonly adopted polytropic relation
\begin{equation}
T_g  \propto
\rho_g^{\gamma - 1},
\label{poly.eq} 
\end{equation}
with $1 \leq \gamma \leq 5/3$.

In the second model we allow a larger parametric freedom and follow the modeling proposed by \cite{vikhlinin2006}. We model the temperature profile using 
\begin{equation}
\label{tpro}
T_g(r)=T_{g,0} \frac{(r/r_t)^{-a}}{[1+(r/r_t)^b]^{c/b}},
\end{equation}
   which describes a power-law declining profile with a transition
   at $ r\sim r_\mathrm{t}$.  In the above model is absent the $t_\mathrm{cool}(r)$ term, 
which accounts for the temperature decline in the central region 
because of radiative cooling (\cite{vikhlinin2006}). 
This is motivated by our choice of the radial range over which the profiles are fitted.

When modeling temperature profiles, a proper treatment would require to fit the model projection along the line of sight against the simulated spectral temperature profiles. Moreover, one has to take into account the presence of different temperature components which can bias the single-temperature fit. Given the large size of our sample, we avoid the involved projection and recovery steps and we fit the model in Eq. (\ref{tpro}) directly to the 3D spectroscopic-like temperature profiles. These are in fact expected to be in good agreement with deprojected temperature profiles derived from spatially resolved spectroscopy. However, in order to explore the crucial bias introduced by spectroscopic temperatures, we additionally apply the same procedure to 3D mass-weighted  temperature profiles, since they provide the azimuthal average of the true gas temperature distribution.

\section{Total mass determination from X-ray observations}
\label{mass.sec}

The total gravitating mass of a galaxy cluster is estimated from X-ray observations assuming that the ICM is in hydrostatic equilibrium and that its temperature and density distributions, as well as the total gravitational potential, are spherical symmetric. These assumptions lead to:
\begin{equation}
M_{\mathrm{}}^{\mathrm{est}}(<r)= - \frac{kT_g(r) \, r}{G \mu m_p} 
\left[ \frac{ d \ln \rho_g(r)}{d \ln r} +
\frac{d \ln T_g(r)}{d \ln r} \right],
\label{mass.eq}
\end{equation}
where $M_{\mathrm{}}^{\mathrm{est}}(<r)$ is the estimated total gravitating mass within a cluster-centric distance $r$, $T_\mathrm{g}(r)$ and $\rho_\mathrm{g}(r)$ are the three-dimensional gas temperature and density profiles at the radius $r$, respectively, and $G$ and $m_\mathrm{p}$ are the gravitational constant and proton mass. We adopt the value $\mu = 0.58$ for the mean molecular weight of the gas.

\subsection{The mass bias}
\label{massbias.sec}

In order to quantify the difference between estimated mass $M^{\mathrm{est}}(<r)$ and the actual mass of the simulated object $M(<r)$, we define the {\it mass bias} as:
\begin{equation}
b(r)=  \frac{M^{est}(<r)-M^{}(<r)}{M^{}(<r)}.
\label{bias.eq}
\end{equation}
The true mass $M^{}(<r)$ is computed by summing the masses of all the particles (dark matter, gas, and star particles) within the radius $r$.

In our observational-like analysis the functions $\rho_\mathrm{g}(r)$ and $T_\mathrm{g}(r)$ in Eq. (\ref{mass.eq}) are modeled as described in Sects. \ref{sx.sec} and \ref{tprof.sec}. The various options for the choice of fitting functions and temperature profiles allow us to investigate and disentangle biases of different origin. In particular we explore three different cases:

\begin{itemize}

\item {\it Polytropic $\beta$-model}: for a given cluster and projection we fit the surface brightness profile and the corresponding projected, spectroscopic-like temperature profile using Eqs. (\ref{1beta-sb.eq}) and (\ref{poly.eq}), respectively. The derived mass bias is referred to as $b_{\mathrm{\beta, T_\mathrm{sl}^\mathrm{2D}}}(r)$.

\item {\it Extended $\beta$-model with spectroscopic temperatures}: for a given cluster and projection we fit the surface brightness profile with Eq. (\ref{rho}). The three-dimensional, spectroscopic-like temperature profile cluster is fitted using Eq. (\ref{tpro}). In this case the mass bias is denoted by $b_{\mathrm{ext \beta, T_\mathrm{sl}^\mathrm{3D}}}(r)$.

\item {\it Extended $\beta$-model with mass-weighted temperatures}: same as the previous case, but the three-dimensional, mass-weighted temperature profile is adopted instead of the spectroscopic-like profile. The derived mass bias is denoted by $b_{\mathrm{ext \beta, T_\mathrm{mw}^\mathrm{3D}}}(r)$.

\end{itemize}

For each of the $\sim 300$ observations the derived mass bias is evaluated at three different overdensities, in correspondence of $\Delta=2500,~500$ and $200$. Cluster-centric distances of $r_{\mathrm{2500}}$ and $r_{\mathrm{500}}$ are usually probed by observations, with $r_{\mathrm{2500}}$ being well inside the largest accessible radius and $r_{\mathrm{500}}$ being typically the largest distance where the ICM temperature can be reliably estimated through X-ray observations \citep[e.g.][]{de02,pi05,vi05,pra07,snowden08,leccardi08}. We additionally choose $r_{\mathrm{200}}$ in order to extend our analysis at radii which will be accessible in the near future \citep[e.g. see][]{reiprich08}, to quantify the effects of extrapolation to large distances and, most important, to investigate the mass bias in dynamically different regions.

We adopt three different values for the outer boundary of the region considered in 
the mass reconstruction by using data (i.e. surface brightness and temperature 
profiles) within $r_\mathrm{up}=0.5,0.75 \,  \mathrm{and} \, 1.05 \times r_\mathrm{200}$ (the inner boundary is $r_\mathrm{low}=0.1 \times r_\mathrm{200}$, see Sects. \ref{sx.sec} and \ref{tprof.sec}). The choice 
of $r_\mathrm{up}$ is of paramount importance to assess the dependence of the mass bias 
upon changes in the gas density slope, which can have a significant impact on mass 
measurement biases. The three different choices allow us to explore a limitation 
that in practice is imposed by the field of view of the detector and the background level of a given exposure.

To summarize, for a given radial range ([0.1-0.5] $r_{\mathrm{200}}$, [0.1-0.75] $r_{\mathrm{200}}$, or [0.1-1.05] $r_{\mathrm{200}}$, with the latter denoted for simplicity by [0.1-1] $r_{\mathrm{200}}$ hereafter), we fit surface brightness and temperature profiles with the analytic functions specific to the adopted model and evaluate Eqs. (\ref{mass.eq}) and (\ref{bias.eq}) at $r_{\mathrm{2500}}$, $r_{\mathrm{500}}$, and $r_{\mathrm{200}}$. It is important to notice that by averaging over the whole sample we find: $<r_{\mathrm{2500}}/r_{\mathrm{200}}>_{\mathrm{sample}}=0.31$ and $<r_{\mathrm{500}}/r_{\mathrm{200}}>_{\mathrm{sample}}=0.67$. As a consequence $M_{\mathrm{}}^{\mathrm{est}}(<r)$, and therefore $b_{\mathrm{}}(r)$, is {\it extrapolated} using the best fit functions at $r_{\mathrm{500}}$ and $r_{\mathrm{200}}$ for the fitting range [0.1-0.5] $r_{\mathrm{200}}$ and at $r_{\mathrm{200}}$ for the fitting range [0.1-0.75] $r_{\mathrm{200}}$. No extrapolation is of course necessary when the fitting range [0.1-1] $r_{\mathrm{200}}$ is adopted.

The large size of the simulated mock sample ($\sim 300$ observations) allows us to use a statistical approach when exploring the dependence of the mass bias on various quantities. For any given mass determination method, overdensity at which the enclosed mass is estimated, radial fitting range, and $\Pi_3$ class $q_\mathrm{i}$ (or whole sample) the mass bias values are modeled using distribution fitting. The data set is always modeled with both normal and Weibull distributions in order to describe symmetric and asymmetric data distributions, respectively. We never find, however, strongly skewed distributions. Even when a Weibull distribution yields a better fit to the data than a normal distribution, its mean and standard deviation are only slightly different from those derived from the normal distribution best-fit. Since these small differences are completely irrelevant for our discussion, we quote only results from normal distribution fitting throughout the paper. From the data set of values $b_\mathrm{i}(r_{\mathrm{\Delta}})$ ($i=\beta, T_\mathrm{sl}^\mathrm{2D}$; $ext \beta, T_\mathrm{sl}^\mathrm{3D}$; $ext \beta, T_\mathrm{mw}^\mathrm{3D}$), distribution fitting yields the mean value $m(b_\mathrm{i}(r_{\mathrm{\Delta}}))$ and standard deviation $\sigma(b_\mathrm{i}(r_{\mathrm{\Delta}}))$. 

Notice that sample S1 is extracted from a volume 8 times larger than the one for S2 (see Sect. \ref{simulations.sec}). This implies that the cluster abundance measured in our sample is not correct. In order to check whether this issue has an important impact on our results we proceed as follows. We create a new sample made of sample S2 and $1/8$ of randomly selected objects from sample S1, for which mean values and standard deviations of the mass bias are computed for the all the cases under consideration. The procedure is repeated many times and the derived values are compared with those computed for the original sample. In all the cases we find that the differences are fully negligible and we therefore report results for the original sample in the reminder of the paper.

Finally, it is important to notice that the four $\Pi_3$ classes $q_\mathrm{i}$ (see Sect. \ref{powerratios.sec}) are defined such that each class contains the same number of objects ($\sim 70$). This guarantees a meaningful comparison between mean values and standard deviations derived from different $\Pi_3$ classes.

\subsection{Results}
\label{betares.sec}
\begin{figure}
\resizebox{\hsize}{!}{\includegraphics{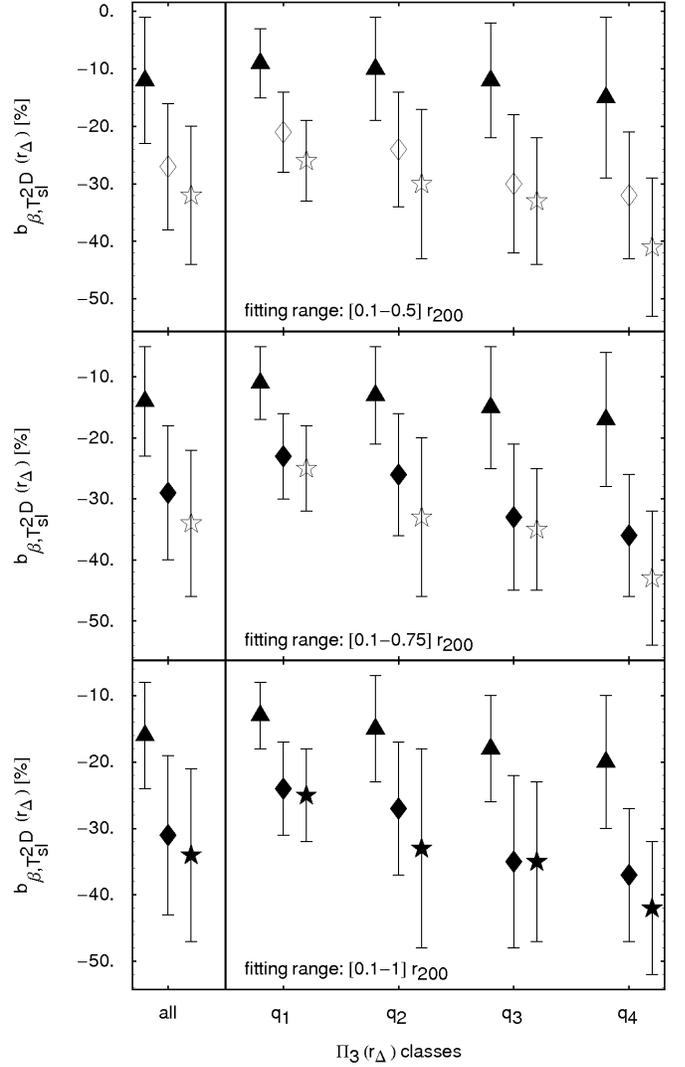}}
\caption{Summary of the results for the mass reconstruction using the polytropic $\beta$-model and projected, spectroscopic-like temperature profiles. The three panels show quantities derived from modeling of the profiles in three different radial ranges. In each panel we show the mean (points) and standard deviation (errorbars) of the mass bias distribution at $r_{\mathrm{200}}$ (stars), $r_{\mathrm{500}}$ (diamonds), and $r_{\mathrm{2500}}$ (triangles). Open/filled symbols indicate quantities derived with/without {\it extrapolation} of the mass profile. The results from distribution fitting are shown for the whole sample (all) and when the four $\Pi_3$ classes are used as sub-samples. The results for different overdensities are shifted horizontally to improve clarity. Notice that the mass bias is shown here in percent.} 
\label{fig:bbetasumS}
\end{figure}

\begin{figure}
\resizebox{\hsize}{!}{\includegraphics{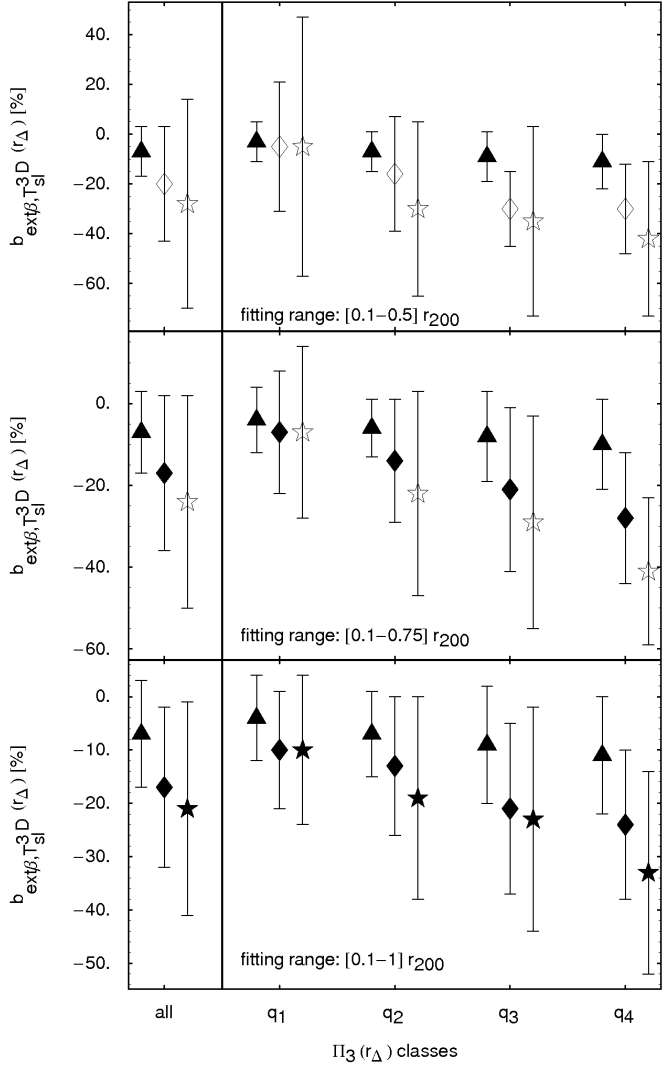}}
\caption{Same as in Fig. \ref{fig:bbetasumS}, but for the mass reconstruction method adopting the extended $\beta$-model and 3D spectroscopic-like temperature profiles. } 
\label{fig:bextbetasumS}
\end{figure}

\begin{figure}
\resizebox{\hsize}{!}{\includegraphics{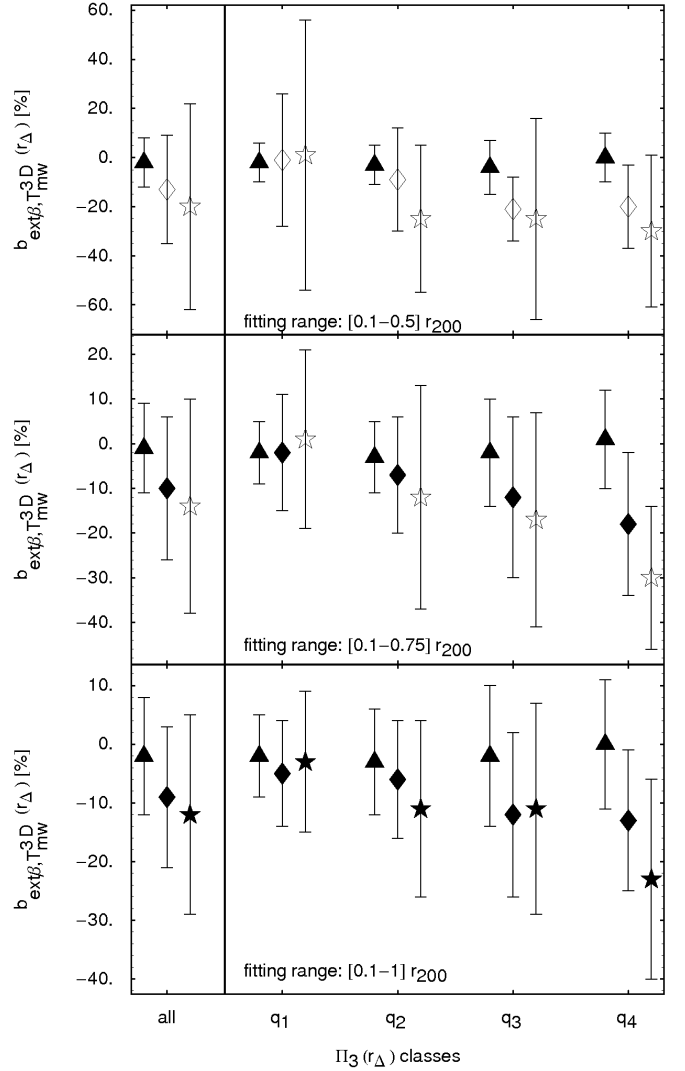}}
\caption{Same as in Fig. \ref{fig:bextbetasumS}, but for results derived using mass-weighted temperature profiles.
} 
\label{fig:bextbetasum}
\end{figure}

Total cluster mass determinations from an X-ray observations, might be in practice affected by more that one source of bias. The number of different cases for which we derive total masses (fitting ranges, best fit functions, temperature definitions, etc.) are of course designed in order to investigate the various sources separately, but also to show the effect of their combination. Our results are reported in Table \ref{tab:bbeta} and conveniently shown in three figures (Figs. \ref{fig:bbetasumS}, \ref{fig:bextbetasumS}, and \ref{fig:bextbetasum}). 
Notice that in order to improve clarity the mean values (data points in the figures) and standard deviations (errorbars in the figures) of the various mass bias distributions are given in percent. The reported values show, in most of the cases, the combined effect of different biases and are very useful to estimate the total bias for a given mass reconstruction method and conditions. Nevertheless, some specific cases and the comparison of mass biases derived under different conditions allow us to quantify the effect of single biases separately.

\begin{table*}[ht]   
\caption{
\label{tab:bbeta} 
Mean values and standard deviations (errors) of the mass bias derived from the observational-like analysis. These are listed in percent for the three adopted models, the three different fitting ranges, and the three overdensities at which they are evaluated. Results are given for the whole sample and for the four $\Pi_3$ classes.}
\centering                            
\begin{tabular}{|c|c|c|ccccc|}        
\hline\hline                 
bias & fitting range & $r_\Delta$ & whole sample & $q_1$ & $q_2$ & $q_3$ & $q_4$ \\ 
\hline        

& & $r_{2500}$ & $-12 \pm 11$ & $-9 \pm 6$ & $-10 \pm 9$ & $-12 \pm 10$ & $-15 \pm 14$  \\  

& $[0.1-0.5] r_{\mathrm{200}}$& $r_{500}$ & $-27 \pm 11$ & $-21 \pm 7$ & $-24 \pm 10$ & $-30 \pm 12$ & $-32 \pm 11$ \\ 

& & $r_{200}$ & $-32 \pm 12$ & $-26 \pm 7$ & $-30 \pm 13$ & $-33 \pm 11$ & $-41 \pm 12$ \\

\cline{2-8}       
                       
& & $r_{2500}$ & $-14 \pm 9$ & $-11 \pm 6$ & $-13 \pm 8$ & $-15 \pm 10$ & $-17 \pm 11$  \\  

$b_{\mathrm{\beta, T_\mathrm{sl}^\mathrm{2D}}}(r_\Delta)$&$[0.1-0.75] r_{\mathrm{200}}$& $r_{500}$ & $-29 \pm 11$ & $-23 \pm 7$ & $-26 \pm 10$ & $-33 \pm 12$ & $-36 \pm 10$ \\ 

& & $r_{200}$ &$-34 \pm 12$ & $-25 \pm 7$ & $-33 \pm 13$ & $-35 \pm 10$ & $-43 \pm 11$ \\ 

\cline{2-8}  
                       
& & $r_{2500}$ & $-16 \pm 8$ & $-13 \pm 5$ & $-15 \pm 8$ & $-18 \pm 8$ & $-20 \pm 10$  \\  

&$[0.1-1] r_{\mathrm{200}}$& $r_{500}$ & $-31 \pm 12$ & $-24 \pm 7$ & $-27 \pm 10$ & $-35 \pm 13$ & $-37 \pm 10$ \\ 

& & $r_{200}$ & $-34 \pm 13$ & $-25 \pm 7$ & $-33 \pm 15$ & $-35 \pm 12$ & $-42 \pm 10$ \\

\hline                       
& & $r_{2500}$ &$ -7 \pm 10$ &$ -3 \pm 8$ &$ -7 \pm 8$ &$ -9 \pm 10$ &$ -11 \pm 11$  \\  

&$[0.1-0.5] r_{\mathrm{200}}$& $r_{500}$ & $ -20 \pm 23$ &$ -5 \pm 26$ &$ -16 \pm 23$ &$ -30 \pm 15$ &$ -30 \pm 18$\\ 

& & $r_{200}$ & $ -28 \pm 42$ &$ -5 \pm 52$ &$ -30 \pm 35$ &$ -35 \pm 38$ &$ -42 \pm 31$ \\ 

\cline{2-8}       
                       
& & $r_{2500}$ &$ -7 \pm 10$ &$ -4 \pm 8$ &$ -6 \pm 7$ &$ -8 \pm 11$ &$ -10 \pm 11$  \\  

$b_{\mathrm{ext \beta, T_\mathrm{sl}^\mathrm{3D}}}(r_\Delta)$&$[0.1-0.75] r_{\mathrm{200}}$& $r_{500}$ & $ -17 \pm 19$ &$ -7 \pm 15$ &$ -14 \pm 15$ &$ -21 \pm 20$ &$ -28 \pm 16$ \\ 

& & $r_{200}$ & $ -24 \pm 26$ &$ -7 \pm 21$ &$ -22 \pm 25$ &$ -29 \pm 26$ &$ -41 \pm 18$ \\ 

\cline{2-8}  
                       
& & $r_{2500}$ & $ -7 \pm 10$ &$ -4 \pm 8$ &$ -7 \pm 8$ &$ -9 \pm 11$ &$ -11 \pm 11$  \\  

& $[0.1-1] r_{\mathrm{200}}$& $r_{500}$ & $-17 \pm 15$ &$ -10 \pm 11$ &$ -13 \pm 13$ &$ -21 \pm 16$ &$ -24 \pm 14$ \\ 

& & $r_{200}$ & $ -21 \pm 20$ &$ -10 \pm 14$ &$ -19 \pm 19$ &$ -23 \pm 21$ &$ -33 \pm 19$ \\

\hline                       
& & $r_{2500}$ & $ -2 \pm 10$ &$ -2 \pm 8$ &$ -3 \pm 8$ &$ -4 \pm 11$ &$ 0 \pm 10$ \\  

&$[0.1-0.5] r_{\mathrm{200}}$& $r_{500}$ & $ -13 \pm 22$ &$ -1 \pm 27$ &$ -9 \pm 21$ &$ -21 \pm 13$ &$ -20 \pm 17$ \\ 

& & $r_{200}$ & $ -20 \pm 42$ &$ 1 \pm 55$ &$ -25 \pm 30$ &$ -25 \pm 41$ &$ -30 \pm 31$ \\ 

\cline{2-8}       
                       
& & $r_{2500}$ & $ -1 \pm 10$ &$ -2 \pm 7$ &$ -3 \pm 8$ &$ -2 \pm 12$ &$ 1 \pm 11$  \\  

$b_{\mathrm{ext \beta, T_\mathrm{mw}^\mathrm{3D}}}(r_\Delta)$&$[0.1-0.75] r_{\mathrm{200}}$& $r_{500}$ & $ -10 \pm 16$ &$ -2 \pm 13$ &$ -7 \pm 13$ &$ -12 \pm 18$ &$ -18 \pm 16$ \\ 

& & $r_{200}$ & $ -14 \pm 24$ &$ 1 \pm 20$ &$ -12 \pm 25$ &$ -17 \pm 24$ &$ -30 \pm 16$ \\ 

\cline{2-8}  
                       
& & $r_{2500}$ & $ -2 \pm 10$ &$ -2 \pm 7$ &$ -3 \pm 9$ &$ -2 \pm 12$ &$ 0 \pm 11$  \\  

& $[0.1-1] r_{\mathrm{200}}$& $r_{500}$ & $ -9 \pm 12$ &$ -5 \pm 9$ &$ -6 \pm 10$ &$ -12 \pm 14$ &$ -13 \pm 12$ \\ 

& & $r_{200}$ & $ -12 \pm 17$ &$ -3 \pm 12$ &$ -11 \pm 15$ &$ -11 \pm 18$ &$ -23 \pm 17$ \\

\hline                           
\end{tabular}
\end{table*}


\subsubsection{The modeling bias}
\label{modbias.sec}
In order to disentangle the bias introduced by inaccurate modeling of temperature and density profiles from other biases, we compare the mean mass bias from polytropic $\beta$ and extended $\beta$-model with spectroscopic temperatures only when no extrapolation of the profiles is needed. Notice that the two models are based on projected and three-dimensional spectroscopic-like temperature profiles, respectively. Nevertheless, the difference between the two types of profiles is small (see Sect. \ref{tprof.sec}) and does not introduce any significant additional bias, in particular for the relaxed clusters.

We therefore focus on the filled symbols in Figs. \ref{fig:bbetasumS} and \ref{fig:bextbetasumS} and the corresponding values in Table \ref{tab:bbeta}. It is important to remark that, as ubiquitously found in simulations, the gas density slope in our simulated clusters considerably steepens with radius. In addition, also radial temperature profiles show a non-trivial radial dependence. Thus, the performance of a given model adopted for the surface brightness and temperature fitting strongly depends on the ability to model these slope changes. 

From a visual inspection of the fitted profiles we find that, as expected, the extended $\beta$-model is extremely accurate in modeling surface brightness and temperature profiles for any given radial fitting range, especially for the most relaxed clusters. The accuracy of this model is reflected in the fact that, when no extrapolation is involved, the bias mean values and standard deviations do not depend on the radial fitting range (compare the filled symbols in the different panels in Fig. \ref{fig:bextbetasumS} and the values in in Table \ref{tab:bbeta}). 

The difference between the mean values derived from extended $\beta$- and polytropic $\beta$-models (compare filled symbols in the bottom panels of Figs. \ref{fig:bbetasumS} and \ref{fig:bextbetasumS} or the corresponding values listed in Table \ref{tab:bbeta}) provides a direct measure of the bias due to the inaccurate modeling provided by polytropic $\beta$-model. The comparison shows that the polytropic $\beta$-model introduces an additional mass underestimate which is on average $\sim 5$, $10$, and $15\%$ at $r_{\mathrm{2500}}$, $r_{\mathrm{500}}$, and $r_{\mathrm{200}}$, respectively. 

\subsubsection{The extrapolation bias}
\label{extrbias.sec}
The bias originating from extrapolating of the mass profiles beyond the radial range probed by observation can be extremely large. While this type of bias can be easily kept under control by simply avoiding any mass estimation/comparison beyond a given outer radius, it may be in practice the cause of many disagreements between different mass estimates (i.e. different X-ray analyses, X-ray and lensing, etc.). 

The bias introduced by extrapolation of course affects all the explored models. However, in order to evaluate its net effect, we restrict our discussion to the results obtained from the extended $\beta$-model and mass-weighted temperature profiles, since spectroscopic temperatures introduce an additional bias that is discussed below. The polytropic $\beta$-model, in addition to being inaccurate (see Sect. \ref{modbias.sec}), also affected by this latter bias. For a given overdensity, the average extrapolation bias can be therefore derived by comparing open with filled symbols in Fig. \ref{fig:bextbetasum} and the corresponding values in Table \ref{tab:bbeta}.

While the extrapolation bias is of course due to the fact that best fit functions for data in a given radial range do not provide a good description of data outside the radial range, it is interesting to notice that it causes a systematic underestimate of the total mass. For the whole sample, the difference between mean mass bias obtained adopting the the fitting range $[0.1-1] r_{\mathrm{200}}$ (i.e. with no extrapolation) and those obtained from the fitting range $[0.1-0.5] r_{\mathrm{200}}$ indicates that extrapolation causes an additional $\sim 10 \%$ average underestimate at $r_{\mathrm{200}}$ (see Table \ref{tab:bbeta}). For a very large sample the average bias introduced by extrapolation is thus not extremely large. Furthermore, if only the most relaxed clusters (i.e., the sub-sample $q_\mathrm{1}$) are taken onto account this bias is fully negligible (see Table \ref{tab:bbeta}). These considerations are true for average bias values. It is however extremely important to notice that standard deviations of mass biases computed from extrapolated values are at least twice as big as those obtained from non-extrapolated values, independently on the dynamical state of the cluster (see the errorbars in Fig. \ref{fig:bextbetasum} and errors in Table \ref{tab:bbeta}). This clearly shows that, for individual objects, extrapolation can lead to extremely large mass over/underestimates even for the most dynamically relaxed objects.  

\subsubsection{The dynamical state bias}
\label{dynbias.sec}
X-ray observations of unrelaxed clusters are expected to yield bias mass estimates because both assumptions of spherical symmetry and hydrostatic equilibrium are not valid. 

Here we quantify the mass bias due to the unrelaxed dynamical state of the cluster by considering: estimates derived from the extended $\beta$-model (because of its accuracy, see \ref{modbias.sec}), 3D mass-weighted temperature profiles (in order to avoid spectroscopic temperature biases), and mass estimates derived without extrapolation (to avoid contamination from extrapolation biases). Our results are shown by the filled symbols in Fig. \ref{fig:bextbetasum} and since for the case under consideration the results do not depend on the adopted fitting range we focus our discussion on the values shown in the lower panel (i.e. the fitting range $[0.1-1] r_{\mathrm{200}}$, see Table \ref{tab:bbeta} for the corresponding values). 

While at $r_{\mathrm{2500}}$ we do not find any appreciable variation of the mean mass bias for the four $\Pi_3$ classes (see Sect. \ref{powerratios.sec} for the definition of these classes), at $r_{\mathrm{500}}$ and $r_{\mathrm{200}}$ we find a clear trend with cluster dynamical state: the mass is on average more underestimated for disturbed clusters than for the relaxed ones. We find, for example, $m(b_\mathrm{ext \beta, T_\mathrm{mw}^\mathrm{3D}}(r_\mathrm{200}))=-3 \%$ and $-23 \%$ for the $q_\mathrm{1}$ class (the most relaxed clusters) and $q_\mathrm{4}$ class (the most disturbed clusters), respectively. 

\begin{figure}
\resizebox{\hsize}{!}{\includegraphics{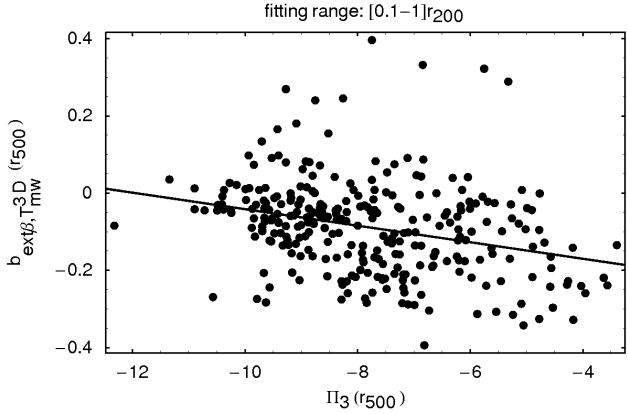}}
\caption{Mass bias as a function of $\Pi_3$ at $r_{\mathrm{500}}$ with estimated masses derived adopting the extended $\beta$-model and 3D mass-weighted temperature profiles. The solid line shows the best fit function $b_{ext \beta, T_\mathrm{mw}^\mathrm{3D}}(r_{\mathrm{500}})=-0.256-0.022 \times \Pi_\mathrm{3}(r_{\mathrm{500}})$.}
\label{fig:fig1}
\end{figure}

Furthermore, we find that at all overdensities the individual mass bias values are more scattered around the mean for disturbed objects than for the relaxed ones, as shown by the standard deviations in Fig. \ref{fig:bextbetasum} and errors in Table \ref{tab:bbeta}. This is illustrated more specifically in Fig. \ref{fig:fig1} for the case of fitting range $[0.1-1] r_{\mathrm{200}}$ and $\Delta=500$. 

These results are particularly relevant because they clearly show that mass underestimates in clusters that are identified as relaxed according to the power ratio method are on average small ($\sim 2-5 \%$), as opposed to the very large bias that affects total mass determinations in disturbed systems. Finally, we remark that, in agreement with other studies \citep[e.g.][]{je07}, we do not find any correlation between $\Pi_3$ and true total mass. 

\subsubsection{The spectroscopic temperature bias}
\label{tempbias.sec}
The difference between emission-weighted, mass-weighted, and spectroscopic ICM temperatures has been addressed by a number of authors \citep[e.g.][]{ma01,gardini04,mazzotta04,rasia05,vikhlinin2006a,valda06}. Special emphasis has been 
given to global temperatures because of their impact on the scaling relations such as the luminosity-temperature and mass-temperature relations. Recent studies indicate that $T_\mathrm{mw}(<r_\mathrm{\Delta}) < T_\mathrm{sl}(<r_\mathrm{\Delta}) < T_\mathrm{ew}(<r_\mathrm{\Delta}) $, where usually $\Delta=500$ \citep[e.g.][]{nagai07,ciotti08}. This relation is also valid for our simulated sample. By averaging over the whole sample we find the ratios: $T_\mathrm{mw}(<r_\mathrm{500}) : T_\mathrm{sl}(<r_\mathrm{500}) : T_\mathrm{ew}(<r_\mathrm{500})=1:1.13:1.27$. In addition we computed average temperatures from the same sample but without applying the masking procedure described in Sect. \ref{sec:map} and find: $T_\mathrm{mw}(<r_\mathrm{500}) : T_\mathrm{sl}(<r_\mathrm{500}) : T_\mathrm{ew}(<r_\mathrm{500})=1:1.08:1.21$. Considering that mass-weighted global temperatures are fully unaffected by masking, these results show that both emission-weighted and spectroscopic-like global  temperatures are on average biased high by $\sim 5 \%$ if cold ICM clumps are not masked out appropriately. 

In the following we focus on the difference between $T_\mathrm{mw}$ and $T_\mathrm{sl}$ since emission weighted temperatures are not relevant in our analysis. As noted by \cite{nagai07} global spectroscopic temperatures within large apertures are higher that mass-weighted ones because they are dominated by the inner, hotter cluster region. It is important to notice that accurate total mass determination methods such as the one adopted in this paper rely on spatially resolved temperature determinations and not on global temperatures. Therefore it is crucial to compare $T_\mathrm{mw}$ and $T_\mathrm{sl}$ radial profiles and how any difference between the two propagates into total mass estimates. 

We compared spectroscopic like and mass-weighted temperature profiles and find, in agreement with \cite{ameglio07}, that $T_\mathrm{mw}(r) > T_\mathrm{sl}(r)$. This inequality, and the fact that the opposite is true when considering global temperatures, can be understood as follows. Global temperatures are computed by averaging within large cluster-centric distances whereas profiles are computed by averaging within radial shells. The density and temperature distributions in the two cases are dramatically different. In the former case the distribution is dominated by the central, hot and dense gas, while in a given radial shell the ICM is obviously distributed in a much narrower range of densities and temperatures. The ICM in our simulations is locally multiphase and made of a dominant component and mostly lower temperature particles which do not substantially contribute to mass-weighted temperature, but bias spectroscopic temperatures low. A simple illustration of the spectroscopic temperature bias for a two-phase ICM can be found in \cite{jia08}. 

It is very relevant to remark that in the computation of both $T_\mathrm{mw}(r)$ and $T_\mathrm{sl}(r)$ all cold clumps are excluded. This implies that local spectroscopic-like temperatures can underestimate mass-weighted ones also because of cold particles which are not part of clumps. Interestingly, we find that our mass reconstruction method depends very weakly on cold clumps removal. Total masses are in fact derived by fitting brightness and temperature profiles with smooth functions of radius, which are in almost all the cases insensitive to the irregularities (spikes or bumps in the brightness profiles and dips or depressions in the temperature profiles) arising from cold clumps when no masking is applied. 
 
In relaxed clusters $T_\mathrm{mw}$ and $T_\mathrm{sl}$ agree very well at small and intermediate radii. The agreement becomes gradually worse with increasingly disturbed dynamical state (as measured by the power-ratio method). This reflects the more complex thermal structure of disturbed clusters, where the fraction of cold gas particles stripped from cold sub-structures is relatively high. Furthermore, we find that the relative difference between $T_\mathrm{mw}(r)$ and $T_\mathrm{sl}(r)$ increases with radius. This is expected because of the weighting scheme used in the computation  of $T_\mathrm{sl}$ and the presence of cold infalling gas in the cluster outskirts. A similar finding is reported in \cite{rasia06}. Finally, the relative difference between $T_\mathrm{mw}$ and $T_\mathrm{sl}$ is higher in more massive clusters than in low mass systems because in massive objects the spread in temperature is larger \citep[e.g.][]{valda06,ameglio07}.   

The underestimate of the true ICM temperature profile introduced by spectroscopic measurement leads to total mass underestimates because hydrostatic masses (Eq. (\ref{mass.eq})) depend on both temperature and its logarithmic derivative. In order to disentangle this effect from the other observational biases discussed in the previous subsections, we consider results obtained from the two extended $\beta$-models and compare results obtained from the one adopting spectroscopic-like profiles with those derived using mass-weighted temperature profiles. The net mass bias introduced by spectroscopic temperatures can be readily estimated by computing the difference between the mean bias values listed in Table \ref{tab:bbeta}. However, a clearer determination can be performed by defining the spectroscopic bias:

\begin{equation}
b_{spec}(r)=  \frac{M^{est}_{ext \beta, T_{sl}^{3D}}(<r)-M^{est}_{ext \beta, T_{mw}^{3D}}(<r)}{M^{est}_{ext \beta, T_{mw}^{3D}}(<r)}.
\label{bspec.eq}
\end{equation}
where $M^{est}_{ext \beta, T_{sl}^{3D}}(<r)$ and $M^{est}_{ext \beta, T_{mw}^{3D}}(<r)$ are total masses estimated through the extended $\beta$-model with spectroscopic-like and mass-weighted temperatures profiles, respectively. Both are derived adopting the the fitting range $[0.1-1] r_{\mathrm{200}}$ in order to avoid any contamination introduced by the extrapolation. 

\begin{figure}
\resizebox{\hsize}{!}{\includegraphics{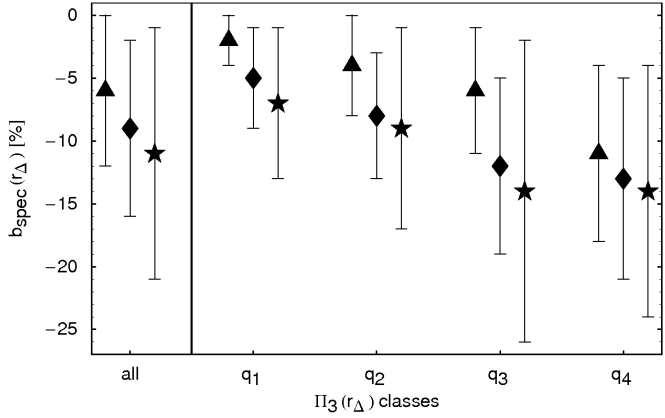}}
\caption{Mean (points) and standard deviation (errorbars) of the spectroscopic bias $b_\mathrm{spec}(r)$ (Eq. (\ref{bspec.eq})) at $r_{\mathrm{200}}$ (stars), $r_{\mathrm{500}}$ (diamonds), and $r_{\mathrm{2500}}$ (triangles). The results are shown for the whole sample (all) and when the four $\Pi_3$ classes are used as sub-samples. The results for different overdensities are shifted horizontally to improve clarity.} 
\label{fig:bspec}
\end{figure}

In Table \ref{tab:bspec} we list the spectroscopic bias mean values and standard deviations computed for the whole sample and for the four $\Pi_3$ classes separately. These values are also shown in Fig. \ref{fig:bspec} for the sake of clarity. The spectroscopic bias is systematic and of the order of $-10 \%$ if the whole sample is considered. However, if the same comparison is performed for the four $\Pi_3$ classes separately one finds that it is larger (in absolute value) in disturbed clusters than in relaxed ones. The mean spectroscopic bias is also more important at larger cluster-centric distances than at small ones. These results are of course directly related to the relative difference between the $T_\mathrm{mw}$ and $T_\mathrm{sl}$ profiles discussed above. 

To reiterate, a crucial point of out findings is that also after the removal of all the resolved ICM cold clumps in the simulations, spectroscopic temperature profiles underestimate the true ones, and so the total mass, because of a diffuse distribution of cold gas particles.

\begin{table*}[ht]   
\caption{
\label{tab:bspec} 
Mean values and standard deviations (errors) of the spectroscopic bias $b_\mathrm{spec}(r_\mathrm{\Delta})$ (Eq. (\ref{bspec.eq})). Values are in percent and listed for three different true mass ranges used in their derivation: all masses (the whole sample), {\it lowM}-sample (clusters with $M(<r_\mathrm{200}) < 1.8 \times  10^{14} \, h^{-1 }M_{\odot}$), and {\it highM}-sample (clusters with $M(<r_\mathrm{200}) > 1.8 \times  10^{14} \, h^{-1 }M_{\odot}$). In each case results are given, at three radii, for the whole sample and for the four $\Pi_3$ classes.}
\centering                            
\begin{tabular}{|c|c|ccccc|}        
\hline\hline                 
mass range & $r_\Delta$ & whole sample & $q_1$ & $q_2$ & $q_3$ & $q_4$ \\ 
\hline        

& $r_{2500}$ &$ -6 \pm 6$ &$ -2 \pm 2$ &$ -4 \pm 4$ &$ -6 \pm 5$ &$ -11 \pm 7$  \\  

all masses & $r_{500}$ &$ -9 \pm 7$ &$ -5 \pm 4$ &$ -8 \pm 5$ &$ -12 \pm 7$ &$ -13 \pm 8$ \\ 

& $r_{200}$ &$ -11 \pm 10$ &$ -7 \pm 6$ &$ -9 \pm 8$ &$ -14 \pm 12$ &$ -14 \pm 10$ \\ 

\hline       
                       
& $r_{2500}$ &$ -4 \pm 4$ &$ -1 \pm 1$ &$ -3 \pm 3$ &$ -6 \pm 4$ &$ -9 \pm 5$  \\

{\it lowM}-sample & $r_{500}$ & $ -7 \pm 4$ &$ -4 \pm 2$ &$ -5 \pm 3$ &$ -9 \pm 4$ &$ -9 \pm 4$ \\ 

& $r_{200}$ &$ -5 \pm 5$ &$ -3 \pm 2$ &$ -4 \pm 4$ &$ -7 \pm 5$ &$ -9 \pm 5$ \\ 

\hline  
                       
& $r_{2500}$ &$ -7 \pm 7$ &$ -2 \pm 1$ &$ -4 \pm 4$ &$ -7 \pm 5$ &$ -12 \pm 9$  \\  

{\it highM}-sample & $r_{500}$ &$ -12 \pm 8$ &$ -7 \pm 5$ &$ -11 \pm 6$ &$ -14 \pm 8$ &$ -16 \pm 9$ \\ 

& $r_{200}$ &$ -17 \pm 11$ &$ -12 \pm 6$ &$ -17 \pm 7$ &$ -19 \pm 14$ &$ -18 \pm 11$ \\

\hline                           
\end{tabular}
\end{table*}

\subsubsection{Mass dependence}
\label{massdep.sec}
Our results concerning the influence of the various observational biases allows us to focus on some relevant cases when exploring the dependence of the mass bias on the total, true mass. We can in fact avoid any contamination from biases such as the modeling bias or the extrapolation bias by considering results provided by extended $\beta$-models and without extrapolation of the mass profiles. 

\begin{figure}
\resizebox{\hsize}{!}{\includegraphics{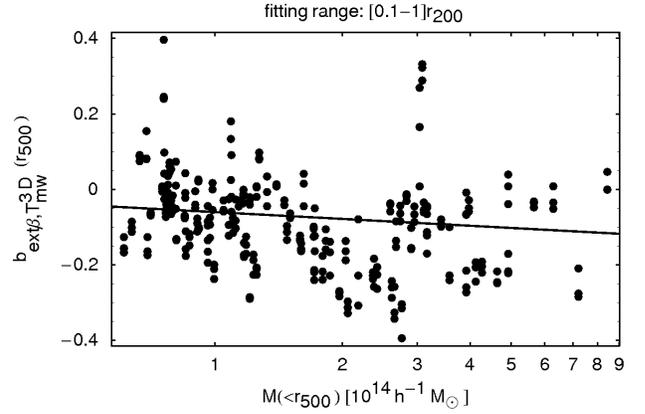}}
\caption{Mass bias from the extended $\beta$-modeling versus total true mass at $\Delta=500$ for the radial fitting range $[0.1-1] r_{\mathrm{200}}$. The estimated total mass is derived using mass-weighted temperature profiles. The solid line shows the best fit function $b_{ext \beta, T_{mw}^{3D}}(r_{\mathrm{500}})=0.78-0.06 \,  \mathrm{log} [M(< r_{\mathrm{500}})/ h^{-1} \, M_{\odot}]$.} 
\label{fig:bvsMextbetar500}
\end{figure}

\begin{figure}
\resizebox{\hsize}{!}{\includegraphics{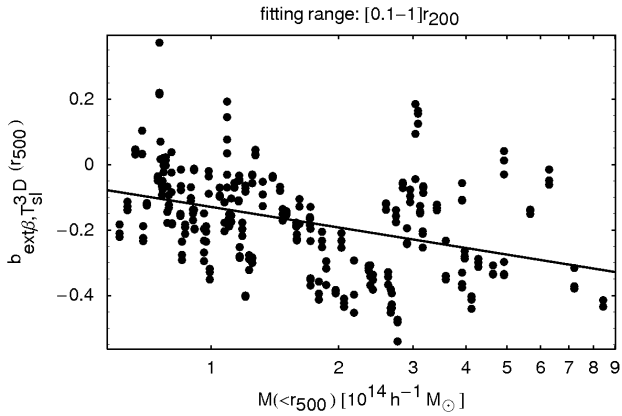}}
\caption{Same as Fig. \ref{fig:bvsMextbetar500}, but with the estimated total mass derived using spectroscopic-like temperature profiles. The solid line shows the best fit function $b_{ext \beta, T_{sl}^{3D}}(r_{\mathrm{500}})=2.79-0.21 \,  \mathrm{log} [M(< r_{\mathrm{500}})/ h^{-1} \, M_{\odot}]$.} 
\label{fig:bvsMextbetar500S}
\end{figure}

As expected, no mass dependence is found for mass biases derived when using the extended $\beta$-model, as shown in Fig. \ref{fig:bvsMextbetar500} in the case of fitting range $[0.1-1] r_{\mathrm{200}}$ and $\Delta=500$. The independence of bias on total mass is found for all fitting ranges and all overdensities. This is, however, only found when mass-weighted temperature profiles are used. Spectroscopic-like temperatures introduce a mass dependence: the average bias is more negative for massive systems, as shown in Fig. \ref{fig:bvsMextbetar500S} for a specific case. This mass dependence arises from the weighting scheme used in the computation of spectroscopic-like temperatures, which is sensitive the cooler part of the gas distribution, and the scale dependency introduced  by cooling on the amount of this cold component versus the cluster mass (see Sect. \ref{tempbias.sec}). 

In order to investigate the dependence of the spectroscopic bias on true mass and at the same time on dynamical state, we compute mean values and standard deviations by considering objects in a given (true) mass interval and $\Pi_3$ class. In order to obtain statistically significant values (i.e. a sufficient number of object in each class), we must adopt a coarse mass binning. After experimenting with different types of mass binning we find that the discussion can be greatly simplified by reporting results for only two sub-samples, which are obtained by ordering the sample by true cluster mass and halving it. In the following these two sub-samples are referred to as {\it lowM}- and {\it highM}-sample.

\begin{figure}
\resizebox{\hsize}{!}{\includegraphics{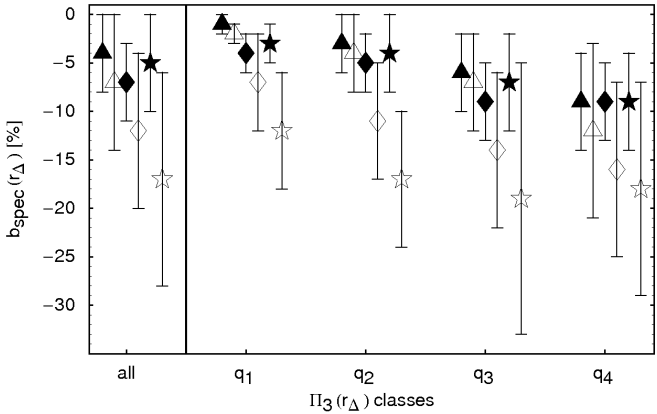}}
\caption{Same as Fig. \ref{fig:bspec}, but for the {\it lowM}- (filled symbols) and {\it highM}-samples (open symbols), separately.} 
\label{fig:bspecM}
\end{figure}

For each of these sub-samples we construct the four $\Pi_3$ classes separately and compute mean and standard deviation of the spectroscopic bias as done above for the whole sample. The results are reported in Table \ref{tab:bspec}, together with those obtained for the whole sample, and shown in Fig. \ref{fig:bspecM}. At $r_{\mathrm{2500}}$ we do not find any significant dependence on mass and therefore the average biases reported in Sect. \ref{tempbias.sec} are valid at any mass. On the other hand, we find a strong mass dependence at $r_{\mathrm{500}}$ and $r_{\mathrm{200}}$ and that its strength increases with cluster-centric distance. For the {\it lowM}-sample the average spectroscopic bias does not significantly vary with overdensity, as opposed to the behavior found for the {\it highM}-sample. Because of the findings reported in Sect. \ref{dynbias.sec}, relaxed clusters ($q_\mathrm{1}$ and $q_\mathrm{2}$ sub-samples) are of particular relevance. The ones in the {\it lowM}-sample are very mildly affected by the spectroscopic bias at all cluster-centric distances, whereas for those in the {\it highM}-sample the mean spectroscopic bias is $\sim -10 \%$ at $r_{\mathrm{500}}$ and $\sim -15 \%$ at $r_{\mathrm{200}}$.
  

\subsubsection{The ideal case}
\label{ideal.sec}
After having examined the various biases affecting the total mass determination, we discuss here the ideal conditions under which the most unbiased mass determination is possible. From our results it is clear that in order to obtain unbiased mass estimates one must select very relaxed clusters (the $q_\mathrm{1}$ sub-sample), adopt a model that can accurately fit the ICM density and temperature profiles (the extended $\beta$-model, here), and avoid extrapolation (notice, however, that the extrapolation bias is on average negligible for relaxed clusters. See Sect. \ref{extrbias.sec}). In addition let us assume that the true thermal structure of the ICM (i.e. $T_\mathrm{mw}(r)$) can be measured. In this case the bias does not depend on the true mass of the system and, at all radii, masses are on average very well reconstructed (see Table \ref{tab:bbeta} and Fig. \ref{fig:bextbetasum}). Notice that there is a residual level of underestimation which might be originating from the still imperfect gas density and temperature modeling or by the departure of the gas from the hydrostatic equilibrium. 

It is of paramount importance, however, to take into account that the true ICM temperature is not accessible through X-ray observations and that we must deal with spectroscopic temperature measurements. As shown in Sect. \ref{tempbias.sec}, the most relaxed clusters are also the less affected by the spectroscopic bias (see Table \ref{tab:bspec} and Fig. \ref{fig:bspec}). As we have shown in Sect. \ref{massdep.sec}, the spectroscopic bias depends on cluster mass. It is negligible in low mass clusters, but important for massive ones especially at large radii. 

\section{Total mass determination from the three-dimensional gas distribution}
\label{masssim.sec}
As shown in Sect. \ref{mass.sec} mass determinations from X-ray observations suffer from various biases. Of course, in addition to these also the assumption of hydrostatic equilibrium (i.e. Eq. (\ref{mass.eq})) plays a crucial role in the total mass reconstruction. Moreover, the derivation of Eq. 
(\ref{mass.eq}) implicitly assumes also spherical symmetry. 
 Whereas the validity of this latter assumption 
can be fairly well tested though an X-ray morphological analysis, the robustness of 
the assumption of hydrostatic equilibrium can not be directly investigated observationally.

Here, in order to investigate the validity of both assumptions we avoid any bias due to the X-ray reconstruction method and make use of the full three-dimensional gas distribution in the simulation. For consistency with the analysis presented in Sec. \ref{mass.sec} we consider only objects with  $T_{\mathrm{sl}}(<r_\mathrm{200}) \geq 2$ keV. 

\subsection{Mass estimators}
\label{massdet.sec}
We introduce the {\it hydrostatic equilibrium mass} estimate, $M_{HE}$ by assuming spherical symmetry and rewriting Eq. (\ref{mass.eq}) as 
\begin{equation}
\label{mhe.eq}
M_{HE}(<r)=-\frac{r^2}{G \rho_g} < \nabla_r P_g(r) >,
\end{equation}
where $P_\mathrm{g}$ is the gas pressure.
As a means to avoid any systematic effect generated by estimating gradients from averaged profiles and to achieve maximum accuracy, we evaluate the term $< \nabla_r P_g(r) >$ directly from the gas particle distribution. To this end, we introduce a spherical shell at the test radius $r$ and a corresponding  set of $40 \times 40 $ grid points with 
angular coordinates uniformly spaced in $cos\theta, \phi$. At the grid point $\vec x_{ gr} $ the pressure gradient is expressed as 
\begin{equation}
\label{grp.eq}
\vec \nabla P_g(\vec x_{gr})=(\gamma-1)\sum m_i u_i \vec \nabla 
W(|\vec x_i -\vec x_{gr}|,h_{\vec g}),
\end{equation}
where $u_i$ is the specific particle internal energy and $\gamma=5/3$ .
The radial component $\nabla_r$ is then extracted by transforming the gradient 
vector according to the grid coordinates and finally $ < \nabla_r P_\mathrm{g}(r) >$ is 
obtained by averaging (\ref{grp.eq}) over the set of grid points. We checked the robustness of the estimated average gradients 
$ < \nabla_r P_\mathrm{g}(r) >$ with a two-fold increase in the number of grid points 
(i.e. $80 \times 80$ ) and found the results to be unaffected by this 
choice.

This procedure allow us to define a mass estimate $M_{HE}$ which is influenced by the cluster dynamical state, but in contrast to the mass determinations discussed in Sect. \ref{mass.sec} it is not affected by any bias originating from the gas density and temperature profiles reconstruction and modeling.

Similarly to Eq. (\ref{bias.eq}) we define the mass bias:
\begin{equation}
b_{HE}(r)=  \frac{M_{HE}(<r)-M^{}(<r)}{M^{}(<r)}.
\label{biasHE.eq}
\end{equation}
In addition, we also consider the {\it hydrodynamical equilibrium mass} estimate \citep{ras04,kay04}:
\begin{eqnarray}
\label{jnsb}
&M_{DE}(<r)&=M_{HE}(<r)+M_{\sigma}(<r)= \nonumber \\
                   &=&M_{HE}(<r) - \frac{\sigma_r^2 r}{G}\left(
\frac {d~ ln~ \rho_g} {d~ln~ r} + \frac {d~ ln ~\sigma_r^2}{d~ ln~ r} +
2 \beta_{\nu}(r)\right ),
\end{eqnarray}
where $\sigma_r$ and $\sigma_t$ are the radial and tangential gas velocity dispersion, respectively, and  $\beta_{\nu}(r)=1 -{\sigma_t^2}/{2\sigma_r^2}$ is the gas velocity anisotropy parameter. The  estimate of the mass term $M_{\sigma}(<r)$ is subject to 
uncertainties because of the irregularities which are present in the 
gas velocity dispersion profiles of the simulated clusters. 
In order to properly estimate the radial derivatives in Eq. \ref{jnsb}
we have therefore regularized the measured profiles by applying 
to them a Savitzky-Golay filter \citep[e.g.][Sect. 14.2]{pre02}. This smoothing procedure is effective in removing the small-scale noise and
yields mass estimates which are much more regular against the radial dependency. As for  $M_{HE}$ we define the mass bias:
\begin{equation}
b_{DE}(r)=  \frac{M_{DE}(<r)-M^{}(<r)}{M^{}(<r)}.
\label{biasDE.eq}
\end{equation}
Once the mass profiles $M_{HE}(<r)$ and $M_{DE}(<r)$ for the $\sim 100$ objects in our sample are computed, we evaluate the mass biases $b_{HE}(r)$ and $b_{DE}(r)$ at $r_{\mathrm{2500}}$, $r_{\mathrm{500}}$, and $r_{\mathrm{200}}$. Finally, mean values and standard deviations of the resulting mass bias distributions are computed. 

Notice that Eq. (\ref{jnsb}) is derived from the Boltzmann equation \citep[e.g.][Chapter 4]{binney1987} by assuming steady-state hydrodynamic equilibrium and taking into account the terms from the gas isotropic pressure and anisotropic velocity dispersion. If the latter are neglected (i.e. it is assumed that the velocity dispersion of the gas is much smaller that its temperature), then Eq. (\ref{jnsb}) yields Eq. (\ref{mhe.eq}). In the derivation it is assumed, of course, that the system is spherically symmetric. 

The steady-state assumption implies no net radial streaming motions (no mean laminar flow, i.e. $<v_r>=0$, which implies $\sigma_r^2= <v^2_r>$) and that all time derivatives can be set to zero. Therefore, the applicability of Eq. (\ref{jnsb}) depends on the degree of spherical symmetry of the system and on the condition $<v_r>=0$, while that of Eq. (\ref{mhe.eq}) additionally depends on the condition $M_{\sigma}(<r)/M_{\Delta}(<r) \ll1$. Hence, the hydrodynamical equilibrium equation naturally provides a more accurate mass estimator than the hydrostatic equilibrium.  

\begin{figure}
\resizebox{\hsize}{!}{\includegraphics{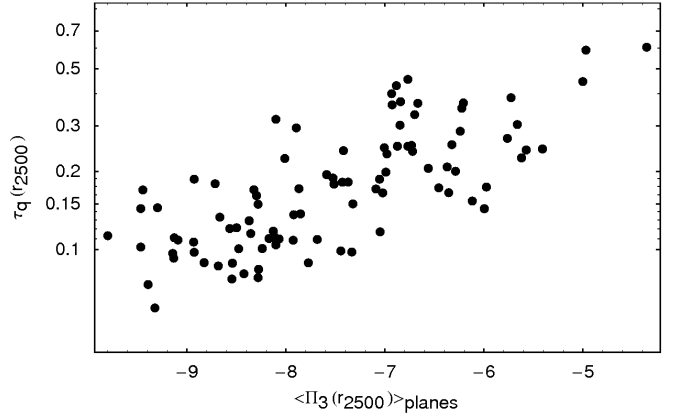}}
\caption{Mean torque parameter $\tau_q$ (which measures the sphericity of the gravitational potential) as a function of $< \Pi_3>_{\mathrm{planes}}$ (which measures the sphericity of the gas distribution) at $r_{\mathrm{2500}}$. For each cluster the value $< \Pi_3>_{\mathrm{planes}}$ is the mean of the three $ \Pi_3$ values computed from three orthogonal surface brightness maps.} 
\label{fig:taupi3}
\end{figure}

The accuracy of spherical symmetry can be assessed by measuring the mean tangential acceleration at the surface of the spherical shell of radius $r$. To this end, in correspondence of each of the grid points at which we evaluate the pressure gradient in Eq. (\ref{grp.eq}), we also evaluate the gravitation acceleration $\vec a_{g}$ of test particle. We then define a mean torque parameter $\tau_q$ by averaging the acceleration components $a_{\theta }^2+a_{\phi}^2$ and $a^2_{r}$ over the set of grid points:

\begin{equation}
\label{tauq}
\tau_q^2= \frac{1}{2} \frac{\sum_g a_{\theta}^2+a_{\phi}^2}{\sum_g a^2_r}.
\end{equation}
The constraint $\tau_q \ll1$  can be used as a condition to validate the spherical symmetry approximation. Accordingly,  the quadrupole 
moment terms can be neglected in the radial component of the total potential gradient and to the lowest order we have: 
\begin{equation}
\label{grm}
  \vec \nabla_{r} \Phi(\vec r ) \simeq \frac{G M(<r)}{r^2} ~~~~~~\tau_q \ll1.
\end{equation}
It is interesting to notice that the degree of spherical symmetry measured directly from the full 3D information provided by the simulations is highly correlated with that measured from X-ray brightness maps. As shown in Fig. \ref{fig:taupi3} for the specific case of $\Delta=2500$ the torque parameter $\tau_q$ and $< \Pi_3>_{\mathrm{planes}}$ are correlated. Here $< \Pi_3>_{\mathrm{planes}}$ is the mean of the three $ \Pi_3$ values computed from three orthogonal surface brightness maps (see Sect. \ref{powerratios.sec}). From a Spearman rank correlation coefficient analysis we find that the correlation is significant at more than the $95 \%$ level at $r_{\mathrm{2500}}$, $r_{\mathrm{500}}$, and $r_{\mathrm{200}}$. Clusters with small values of torque parameter $\tau_q$, i.e. whose potential has a high degree of sphericity, are therefore identified as spherical systems when selected by means of the power ratio method.   

The level of radial streaming motions at the radius $r$ is directly quantified through the radial Mach number $<v_r>/c_s$, where the gas sound speed $c_s$, is computed from $c_s^2(r)=5 k T_\mathrm{g}(r)/(3 \mu m_\mathrm{p}) $

It follows that the applicability of Eqs. (\ref{mhe.eq}) and (\ref{jnsb}) to measure the cluster gravitating mass within the test radius $r$ is then limited to those systems for which at the radius $r$ the conditions $\tau_q\ll1$ and $|<v_r>/c_s| \ll1$ are satisfied.

\subsection{Results}
\label{mass3dres.sec}
\begin{figure}
\resizebox{\hsize}{!}{\includegraphics{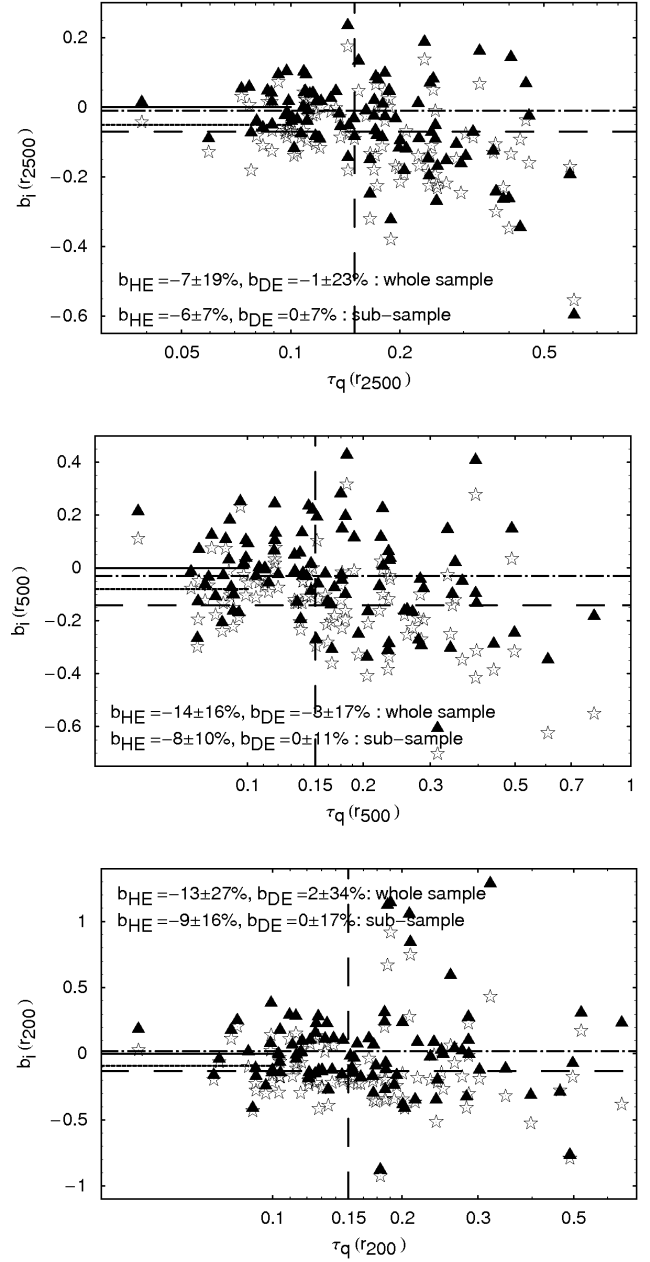}}
\caption{Mass biases $b_{HE}$ (open stars) and $b_{DE}$ (filled triangles) as a function of mean torque parameter $\tau_q$ at $r_{\mathrm{2500}}$, $r_{\mathrm{500}}$, and $r_{\mathrm{200}}$. In each panel we show the mean values derived from the whole sample for $b_{HE}$ (dashed line) and $b_{DE}$ (dot-dashed line), and the mean values (plotted only up to 0.1) derived from the sub-sample with $\tau_q < 0.15$ and $|<v_r>/c_s| < 0.1$ for $b_{HE}$ (short-dashed line) and $b_{DE}$ (solid line). In each panel are also reported these mean values along with the standard deviations. Notice that in order to focus on the bulk of the data we have omitted the results for a few clusters with very large bias and large $\tau_q $.}
\label{fig:bitauq}
\end{figure}

\begin{figure}
\resizebox{\hsize}{!}{\includegraphics{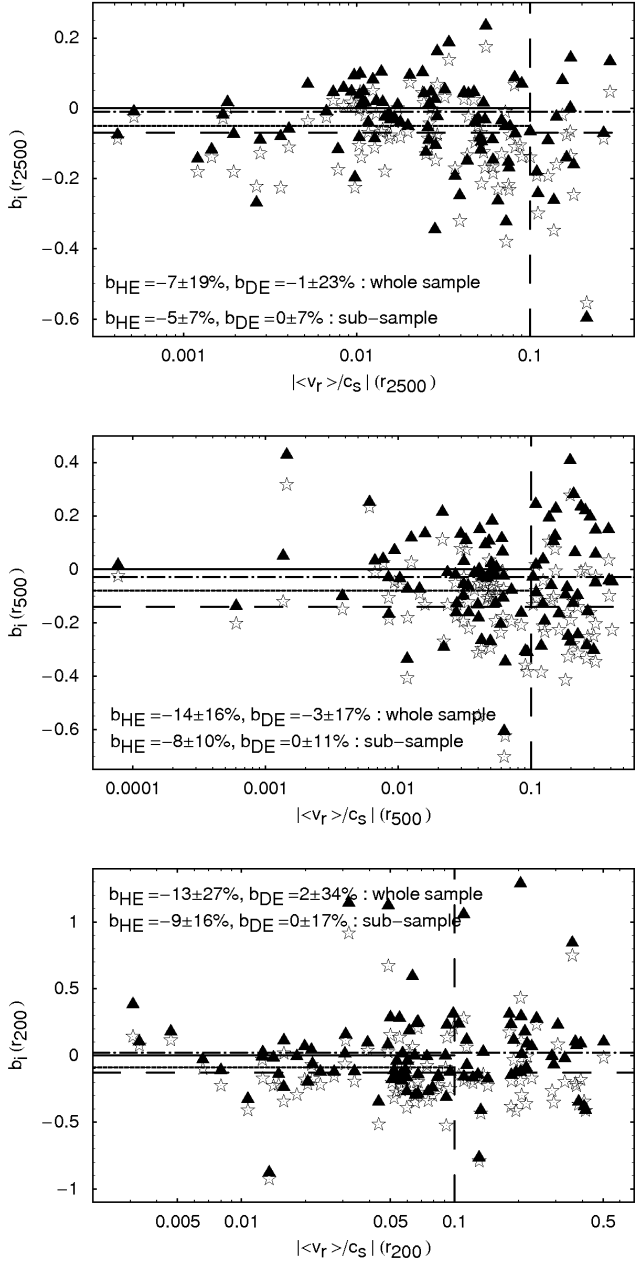}}
\caption{Mass biases as a function of  radial Mach number $<v_r>/c_s$. Symbols, lines, and values are as explained in Fig. \ref{fig:bitauq}. Also here few clusters with very large bias and large $|<v_r>/c_s| $ are omitted from the plot to improve clarity.}
\label{fig:bivr}
\end{figure}

The dependence of the mass biases $b_{HE}$ and $b_{DE}$ on the mean torque parameter $\tau_q$ and the radial Mach number $<v_r>/c_s$  are shown in Figs. \ref{fig:bitauq} and \ref{fig:bivr}, respectively. The figures show that, for a given overdensity, we measure large mass biases (both positive and negative) for clusters with large $\tau_q$ and $|<v_r>/c_s|$. As we move from large to small values of the parameters (i.e. to an increasing degree of sphericity of the cluster or to small radial streaming velocities) the scatter of the measured biases substantially decreases. These results demonstrate that, in addition to sphericity, the radial Mach number $<v_r>/c_s$ is a key parameter in the study of the reliability of the hydrostatic and hydrodynamical mass estimators.


\begin{figure}
\resizebox{\hsize}{!}{\includegraphics{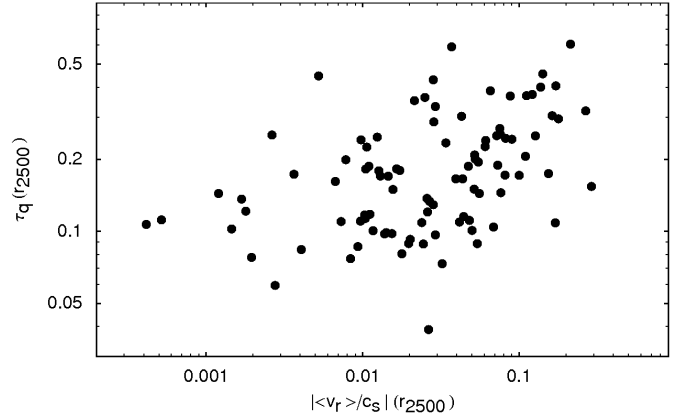}}
\caption{Mean torque parameter $\tau_q$ as a function of Mach number $|<v_r>/c_s|$ at $r_{\mathrm{2500}}$.} 
\label{fig:tauvr}
\end{figure}

It is important to notice that the parameters $\tau_q$ and $<v_r>/c_s$ are not uncorrelated. We find in fact that objects with large $\tau_q$ also tend to have large $|<v_r>/c_s|$, as shown in Fig. \ref{fig:tauvr} for $\Delta=2500$. The correlation between these parameters is found to be significant at more than the $95 \%$ level at all the three considered overdensities.

In addition to the considerable decrease of the scatter with decreasing values of $\tau_q$ and $|<v_r>/c_s|$, a visual inspection of Figs. \ref{fig:bitauq} and \ref{fig:bivr} indicates that on average the mass estimators $M_{HE}$ is biased low and that masses computed through the more accurate estimator $M_{DE}$ are on average well recovered. This is quantitatively shown by mean values derived from the whole sample which are listed in Table \ref{tab:bHEDE}. 

It is crucial to remark that average biases derived from the hydrostatic mass estimator $M_{HE}$ are in agreement with those obtained from the X-ray mass reconstruction procedure presented in Sect. \ref{mass.sec}. In fact, a comparison between the values in Table \ref{tab:bHEDE} and those listed in Table \ref{tab:bbeta} for extended $\beta$-model with mass-weighted temperature profiles (since these follow the virial relationship more accurately that spectroscopic ones) shows that in the two cases the true mass is similarly underestimated. Furthermore, we find that, as for the mass bias derived using mass-weighted temperature in Sect. \ref{mass.sec}, both $b_{HE}$ and $b_{DE}$ do not depend on cluster true mass.

Most important, the mean values of the $b_{HE}$ and $b_{DE}$ distributions show that if in addition to the thermal pressure also the ICM non-thermal pressure component (i.e. the gas anisotropic velocity dispersion) is taken into account, the total mass of the system is on average better reconstructed \citep[see also][]{ras04,kay04}.

\begin{table}[ht]   
\caption{
\label{tab:bHEDE} 
Mean values and standard deviations (errors) of the mass biases $b_{HE}$ and $b_{DE}$ derived from the whole sample and for the sub-sample with $\tau_q < 0.15$ and $|<v_r>/c_s| < 0.1$ . Values are given in percent at three different overdensities.}
\centering                            
\begin{tabular}{|c|c|cc|}        
\hline\hline                 
bias & $r_\Delta$ & whole sample & sub-sample \\ 
 & &  &  ($\tau_q < 0.15$, $|\frac{<v_r>}{c_s}| < 0.1$)  \\ 
\hline        

& $r_{2500}$ &$ -7 \pm 19$ &$ -5 \pm 7$   \\  

$b_\mathrm{HE}(r_\mathrm{\Delta})$ & $r_{500}$ &$ -14 \pm 16$ &$ -8 \pm 10$ \\ 

& $r_{200}$ &$ -13 \pm 27$ &$ -9 \pm 16$  \\ 

\hline       
                       
& $r_{2500}$ &$ -1 \pm 23$ &$ 0 \pm 7$   \\  

$b_\mathrm{DE}(r_\mathrm{\Delta})$ & $r_{500}$ & $ -3 \pm 17$ &$ 0 \pm 11$  \\ 

& $r_{200}$ &$ 2 \pm 34$ &$ 0 \pm 17$ \\

\hline                           
\end{tabular}
\end{table}


In oder to investigate the accuracy of the mass estimators $M_{HE}$ and $M_{DE}$ more precisely we proceed as follows. At a given overdensity, we create sub-samples of clusters with $\tau_q$ and $|<v_r>/c_s|$ smaller that some given threshold values and in each case we compute mean values and standard deviations of $b_{HE}$ and $b_{DE}$ . It is important to find a compromise between small threshold values for $\tau_q$ and $|<v_r>/c_s|$ (in order to guarantee as much as possible the applicability of the mass estimators) and a sub-sample with enough objects so as to allow a meaningful distribution fitting. We experimented with various cuts and find that the optimal choice is to select clusters with $\tau_q < 0.15$ and $|<v_r>/c_s| < 0.1$, which yields sub-samples of at least $\sim 40$ objects at each overdensity. Results derived from this sub-sample are listed in Table \ref{tab:bHEDE}. We remark that the results are not sensitive to exact choice of the threshold values. As expected the standard deviations are much smaller than those derived for the whole sample. Standard deviation for the sample with $\tau_q > 0.15$ and $|<v_r>/c_s| > 0.1$ are $\sim 2-3$ times larger that those obtained for the sample with $\tau_q < 0.15$ and $|<v_r>/c_s| < 0.1$. 

The analysis of simulated objects for which the applicability criteria of Eqs. (\ref{mhe.eq}) and (\ref{jnsb}) are most valid allows us, therefore, to draw more robust conclusions on the reliability of the mass estimators $M_{HE}$ and $M_{DE}$. Hydrostatic masses are biased low by $\sim 5 \%$ at $r_{\mathrm{2500}}$ and by $\sim 10 \%$ at $r_{\mathrm{500}}$ and  $r_{\mathrm{200}}$, whereas the hydrodynamical equilibrium mass estimator is unbiased at any overdensity.  

\section{Discussion and conclusions}
\label{conclusions.sec}
This work is aimed at elucidating the reliability of X-ray total mass estimates in clusters of galaxies using N-body/SPH simulation of a large sample of clusters. The physical modeling the gas includes radiative cooling, star formation, supernovae heating, and metal enrichment. The large number of simulated clusters enables us to derive very robust conclusions through a statistical analysis of the sample. The total mass is recovered adopting an observational-like approach and compared with the true mass in the simulations. Surface brightness and temperature profiles, that we generate from the simulations, are used to estimate the cluster mass at different overdensities ($r_{\mathrm{2500}}, r_{\mathrm{500}}$, and $r_{\mathrm{200}}$) by means of the hydrostatic equilibrium equation. We explore various models and conditions under which the mass is reconstructed in order to entangle different mass biases. In addition, a power ratio analysis of the surface brightness maps allows us to assess the dependence of the mass bias on cluster dynamical state. Moreover, we perform a study on the reliability of  hydrostatic and hydrodynamical equilibrium mass estimates using the full three-dimensional gas distribution in the simulation.

In the following we list and discuss our main findings.

\begin{enumerate}
\item Our analysis shows that it is very important to use analytical models with a large amount of parametric freedom when modeling the shape of the ICM temperature and surface brightness radial profiles. A model with a low degree of sophistication such as the polytropic $\beta$-model can introduce a very large {\it modeling bias}. Compared to the more sophisticated extended $\beta$-model, which is found to be extremely accurate in following the slope changes of the gas profiles, it additionally leads to average mass underestimates of the order of $\sim 5$, $10$, and $15\%$ at $r_{\mathrm{2500}}$, $r_{\mathrm{500}}$, and $r_{\mathrm{200}}$, respectively.

\item The bias originating from extrapolating of the mass profiles beyond the radial range probed by observation can be extremely large. We find that the underestimate from extrapolation alone is of the order of $\sim 10 \%$ at $r_{\mathrm{200}}$, and lower at smaller cluster-centric distances, when considering an average over the whole sample. However, for individual objects, the {\it extrapolation bias} can be as large as  $\sim 50 \%$. 

\item The unrelaxed {\it dynamical state} of a cluster can also lead to mass underestimates. The total mass is on average biased lower for disturbed clusters than for the relaxed ones. Furthermore, we find that the bias values are much more scattered around the mean for disturbed objects than for the relaxed ones. If mass-weighted temperature profiles are adopted, the mean mass bias is at most $\sim -5 \%$ at all cluster-centric distances for relaxed clusters, but can be as large as $\sim -20 \%$ for the most disturbed ones.
 
Our results are in excellent agreement with the findings of \cite{je07}, as shown by the comparison of Fig. \ref{fig:fig1} in this work and Fig. 9 (top panel) in their paper, where the correlation between the same quantities are shown. This agreement is of particular importance, considering that the analysis by \cite{je07} is based on simulations performed with the adaptive mesh refinement code {\it Enzo}. Our findings are also in good agreement with the results of \cite{kay07}.

\item Mass estimates derived using spectroscopic temperatures are lower that those derived from mass-weighted temperatures (i.e. the true gas temperatures). We find that the {\it spectroscopic temperature bias}, i.e. the bias originating from spectroscopic temperatures alone, is of the order of $-10 \%$ for the whole numerical sample. A similar value is found by \cite{kay07}. 

Mass underestimates derived adopting spectroscopic-like temperature profiles ($7$ and $17 \%$ on average at $\Delta=2500$ and 500 for the whole sample) are in good agreement with the values found by \cite{nagai07} ($12$ and $16 \%$ are the corresponding values), whose results are based on mock X-ray observations derived from simulations performed with an Eulerian code. We also find very good agreement by performing the same comparison for relaxed and unrelaxed clusters separately. We notice that total mass biases derived from the simulations by \cite{nagai07} and adopting mass-weighted profiles are on average $-7 \%$ at $r_{\mathrm{500}}$ for relaxed clusters \citep{lau07}. The corresponding value that we find from our Lagrangian simulations is $-5 \%$. Considering the different nature of the numerical codes and the different implementation of the various physical processes, the agreement is extremely relevant. 

Our results are in tension with the findings of \cite{rasia06}, who find, on average, much stronger biases. We notice, however, that a fair comparison is not possible since the work by \cite{rasia06} focussed only on 5 clusters.  

Our analysis also shows that the spectroscopic bias depends on dynamical state. We find that the spectroscopic bias is rather small ($-2,-5$, and $-7 \%$ at $\Delta= 2500, 500$ and 200) for the most relaxed clusters and of the order of $-12 \%$ for the most disturbed ones. 

 \item While the mass bias derived from mass-weighted temperature profiles does not depend on true cluster mass, the one derived from spectroscopic-like temperature exhibits a strong {\it mass dependence}. For clusters with $M(<r_\mathrm{200}) < 1.8 \times  10^{14} \, h^{-1 }M_{\odot}$ we find that, at all radii, the spectroscopic bias contributes on average to the total bias with less than $7 \%$, and less than $4 \%$ if one considers only the most relaxed clusters. On the other hand, for clusters with $M(<r_\mathrm{200}) >1.8 \times  10^{14} \, h^{-1 }M_{\odot}$ we find that the spectroscopic bias is larger: e.g. $-7 \%$ and $-12 \%$ for the relaxed clusters at $r_{\mathrm{500}}$ and $r_{\mathrm{200}}$, respectively. 

\item Even in the {\it ideal case}, i.e. when the mass of relaxed clusters is estimated without extrapolation and adopting a very accurate model for the ICM density and temperature profiles, total masses are affected by the spectroscopic temperatures. In this case the mean total mass bias is $\sim (-4,-3)$, (-12,-9), and $(-15,-3) \%$ at $r_{\mathrm{2500}}, r_{\mathrm{500}}$, and $r_{\mathrm{200}}$ for the (most, least) massive clusters in our simulated sample. 

\item Even if we assume that the true (mass-weighted) ICM temperature can be reconstructed, masses are biased low (by -2, -9, and $-12 \%$ at $\Delta=2500$, 500, and 200, respectively, when taking the average over the whole sample).

\item The latter finding prompted us to investigate the possible {\it violation of the hydrostatic equilibrium} by using the full, three-dimensional information provided by the simulations. In agreement with the results from our observational-like analysis, we find that the hydrostatic equilibrium assumption yields masses underestimated by $\sim 10-15 \%$. This implies that the origin of the bias is not originated by the X-ray reconstruction method. The same level of mass bias has been found by \cite{ras04} and \cite{burns07}.

\item In order to elucidate the origin of this bias we compute masses using both {\it hydrostatic and dynamical equilibrium} mass estimators. The dynamical equilibrium takes into account for both thermal and non-thermal pressure of the ICM. We find that the mass estimated through dynamical equilibrium are on average well estimated. The mean bias profiles of the "feedback" clusters simulated by \cite{kay04} are in very good agreement with our average values at all overdensities.  

\item We explore the conditions of applicability of both estimators, i.e. spherical symmetry (torque parameter $\tau_q\ll1$) and small radial streaming velocities (radial Mach number $|<v_r>/c_s| \ll1$). We find that the biases $b_{HE}$ and $b_{DE}$ do not depend on cluster true mass and that their scatter decreases with decreasing $\tau_q$ and $|<v_r>/c_s|$.

\item For clusters with small values of $\tau_q < 0.15$ and $|<v_r>/c_s| < 0.1$ we find that the average mass underestimate found from the hydrostatic equilibrium estimator ($5 \%$ at $r_{\mathrm{2500}}$ and $10 \%$ at $r_{\mathrm{500}}$ and $r_{\mathrm{200}}$) is extremely well corrected by adopting the dynamical equilibrium estimator.

\end{enumerate}

Our results of course depend on the physics included in the simulation. Our modeling of the gas physical processes is incomplete and does not include, for example, energy feedback from active galactic nuclei (AGN). The inclusion of more realistic physics is particularly relevant in the cluster cores, since it is supposed to provide additional heating to the gas which could help to solve the cooling flow problem as well as the overcooling of baryons actually present in current simulations \citep[e.g.][]{vo05}.

The effect on the thermal status of the ICM of energy feedback from the AGN are however not expected to modify in a significant way the ICM properties outside the cluster cores. The validity of this assumption is justified from the radial behavior of the measured temperature profiles, which are in good agreement at $ r \gtrsim 0.1 \times r_\mathrm{200}$ with the present simulations \citep{valda06} and the ones of other authors \citep[e.g.][]{nagai07b}.

Importantly, we find that, in addition to cold gas clumps, also the diffuse cold gas component which is left after the removal of all resolved clumps substantially biases spectroscopic temperatures low. The good agreement of our temperature profiles with those found by other authors using different codes suggests that the amount of the cool gas component present in our simulations is not affected by numerical issues,

There is another issue, which is connected to the gas physical modeling
in the simulations, that could be potentially relevant for our analysis. In the runs performed here the artificial viscosity is treated according to the standard 
SPH formulation \citep{mo05}, which is a numerical scheme comparatively viscous \citep{momo97} and inadequate to follow the development of fluid turbulence. It has been shown that the level of kinetic energy in random gas motion could be as high as $ \sim 30 \%$ of the gas thermal energy \citep{dolag05,vazza06}, when the numerical viscosity scheme is generalized as in \cite{momo97} to properly treat fluid turbulence.

It is worth noticing that very high levels of random motions are inconsistent with the upper limits recently derived from the comparison of X-ray and weak lensing mass estimates \citep{mahdavi08,zhang08}. Direct measurements of the various sources of non-thermal pressure (gas motions, cosmic rays, magnetic fields, etc.) are extremely difficult, and the comparison between X-ray and lensing masses presently provides the most significant constraints on the level of non-thermal pressure in the ICM. \cite{zhang08} obtain a ratio of $1.09 \pm 0.08$ between the weak lensing and X-ray mass estimates extracted at $r_\mathrm{500}$ from a sample of 19 clusters. At the same overdensity, \citep{mahdavi08} find that the ratio between X-ray and lensing is $0.78 \pm 0.09$ ($0.85 \pm 0.10$ after correction for excess structure along the line of sight) for a sample of 18 clusters. The fairly good agreement between these recent observational results and the values found in the present analysis and other similar studies therefore implies a low level of turbulence present in the ICM, thereby suggesting that the gas physics outside the cluster cores is well approximated by the simulations presented here.

While we have shown that at present both theoretical models and observations seem to indicate a deviation from the hydrostatic equilibrium of the order of $ \sim 10 \%$, further investigations are needed in order to draw robust conclusions on this important issue. Future measurements of  X-ray and lensing masses for large samples and measurements of the ICM velocity structure will provide valuable information on the validity of the hydrostatic 
equilibrium and therefore on the reliability of X-ray clusters as cosmological probes. 

Moreover, these measurements will also be of particular relevance for constraining
gas physics in galaxy clusters. The amount of physical viscosity is in particular a key 
parameter which quenches the level of turbulence present in the ICM.
Physical viscosity has been incorporated in hydrodynamic simulations of galaxy clusters 
by \cite{si06} and it was shown that even a modest
amount of  physical viscosity has significant consequences on ICM properties.
Therefore X-ray mass estimates, extracted from a statistically meaningful sample of hydrodynamical SPH simulations which include  physical viscosity, could be profitably 
used to indirectly measure the level of ICM viscosity when contrasted against X-ray and weak lensing mass measurements.

These constraints will also likely have a significant impact on those scenarios in which 
the cooling flow problem is solved by providing additional heating to the gas through energy dissipation.

\begin{acknowledgements}

We acknowledge the anonymous referee for detailed and valuable comments that helped in improving the presentation of
the manuscript. We would like to thank Daisuke Nagai and Alexey Vikhlinin for useful comments. RP thanks Stefano Borgani for stimulating discussions.
\end{acknowledgements}     


\end{document}